\documentclass[a4paper]{report}
\usepackage[utf8]{inputenc}
\usepackage[T1]{fontenc}
\usepackage{RJournal}
\usepackage{amsmath,amssymb,array}
\usepackage{booktabs}

\usepackage{multirow, comment}
\usepackage{graphicx}
\graphicspath{ {./figures/} }
\usepackage{subcaption}
\usepackage{float}

\renewcommand\thesection{\arabic{section}}
\renewcommand\thesubsection{\thesection.\arabic{subsection}}
\makeatletter
\@addtoreset{subsection}{section}
\makeatother
\usepackage{titlesec}
\titleformat{\subsection}[block]      
  {\normalfont\bfseries\fontsize{10}{11}\selectfont} 
  {\thesubsection}
  {0.8em}
  {}

\titlespacing*{\subsection}{0pt}{4ex plus .5ex minus .2ex}{2ex plus .5ex}

\begin{document}

\sectionhead{Contributed research article}
\volume{XX}
\volnumber{YY}
\year{20ZZ}
\month{AAAA}

\begin{article}
\title{SensIAT: An R Package for Conducting Sensitivity Analysis of Randomized Trials with Irregular Assessment Times}
\author{by Andrew Redd, Yujing Gao, Bonnie B. Smith, Ravi Varadhan, Andrea Apter, and Daniel Scharfstein}

\maketitle

\abstract{
This paper introduces an R package SensIAT that implements a sensitivity analysis methodology, based on augmented inverse intensity weighting, for randomized trials with irregular and potentially informative assessment times. Targets of inference involve the population mean outcome in each treatment arm as well as the difference in these means (i.e., treatment effect) at specified times after randomization. This methodology is useful in settings where there is concern that study participants are either more, or less, likely to have assessments at times when their outcomes are worse. In such settings, unadjusted estimates can be biased. The methodology allows researchers to see how inferences are impacted by a range of assumptions about the strength and direction of informative timing in each arm, while incorporating flexible semi-parametric modeling.  We describe the functions implemented in SensIAT and illustrate them through an analysis of a synthetic dataset motivated by the HAP2 asthma randomized clinical trial.
}

\section{Introduction}

In many randomized trials, the timing and frequency of outcome assessments varies by participant, either by design or in practice, and may also be related to outcome values.  For example, a downturn in health may lead participants to postpone or miss a data collection appointment or, alternatively, to seek care at an earlier time than planned.  Failure to account for such informative assessment in the analysis can bias inference for treatment effects.  This is a similar---though distinct---problem to trials with informative missing data, where it is recommended to conduct a sensitivity analysis \citep{nationalacademy2010missingdata};  and sensitivity analyses should similarly be incorporated in the analysis of trials with irregular and potentially informative assessment timing.  

Recently, \citet{smith2024sa} developed a sensitivity analysis methodology for this setting.  Their method builds on the inverse intensity weighting (IIW) approach first developed by Lin, Scharfstein, and Rosenheck \citep{lin2004dependent}, which accounts for the impact on assessment times of any variables in the study data that occur prior to the assessment in question.  The method of \citet{smith2024sa} allows researchers to also account for the impact of the current outcome, with a sensitivity parameter that quantifies the strength and direction of this dependence.   Within this framework, the authors derived a new \emph{augmented inverse intensity weighted} (AIIW) estimator for the mean outcome at a given time under each treatment condition.  The AIIW estimators are influence function (IF)-based estimators that incorporate flexible semi-parametric models, yet allow fast root-$n$ rates of convergence thanks to a product bias property \citep{naimi2021machinelearning}.

The new R package \CRANpkg{SensIAT} provides code for implementing this sensitivity analysis methodology.  The package is available from the \href{https://cran.r-project.org/package=SensIAT}{Comprehensive R Archive Network (CRAN)} and  \href{https://github.com/UofUEpiBio/SensIAT}{GitHub} \citep{GHSensIAT}.  This paper describes the functions implemented in \CRANpkg{SensIAT} and illustrates them through an analysis of a synthetic dataset motivated by the HAP2 asthma randomized clinical trial \citep{apter2020patientadvocates}.  The paper is organized as follows.  Section 2 reviews elements of the methodology, with an emphasis on modeling choices that the user will need to make when using the package.  Section 3 discusses the implementation of the package. 
Section 4 illustrates use of the software in a data analysis using synthetic data based on the HAP2 trial.
Section 5 presents the results of a simulation study.  Section 6 discusses some future extensions.

\section{Methodology}\label{sec:methods}

\subsection{Setting and notation}

 The methodology used here is for a two-arm randomized trial ("treatment" versus "control") in which times of outcome assessments vary by participant.  Interest is in the population mean outcome in each arm as well as the difference in these means (i.e., treatment effect) at one or more fixed target times, which we denote by $t^*$.  The method assumes:
 
 \begin{itemize}
\item \underline{Randomized treatment}.

\item \underline{\smash{Irregular assessment times}}:  there is variability in the timing of assessments, with assessments taking place over a range of times around each target time $t^*$.

\item \underline{Continuous outcome}:
the marginal mean model (see Equation \eqref{marg_mean}) uses an identity link function.

\item \underline{\smash{No dropout}}:  no participants drop out of the study.  However, some participants may have fewer assessments than others; and if the study protocol calls for a certain number of assessments, a participant could have more or fewer than this number of assessments.

\item \underline{\smash{Non-future dependence}}:  whether a participant is assessed at a given time is not associated with future values of the outcome, or with past unobserved values of the outcome, except through past observed data and the (possibly unobserved) current value of the outcome.  Note that this assumption may not be met, for example, if participants receive care at an outcome assessment appointment that could impact future outcome values. 

\end{itemize}

For a random individual, let $Y(t)$ be  the (possibly unobserved) outcome at time $t$ and let $N(t)$ be the number of assessments that they have had up through time $t$.  Let $\overline{\boldsymbol{O}}(t)$ denote the individual's \emph{observed past}, that is, all of their study data observed before time $t$, including baseline data, treatment assignment, times of assessments prior to $t$, and all data collected at each assessment prior to $t$.  Let $\Delta N(t)$ be the indicator that the individual had an assessment at time $t$.  

We use the subscripts $i$ and $j$ when we refer to data specific to individuals $i$ and $j$, respectively.  Let $K_i$ be the number of observed assessments for individual $i$ and $\{ T_{ik}: k=1,\ldots, K_i \}$ be the set of their assessment times.

All models are fit separately by treatment arm. We suppress the dependence on treatment arm in the notation for all computations dealing with a single treatment arm.  In each arm, let $dF( y(t) \mid \Delta N(t)=1, \overline{\boldsymbol{O}}(t) )$ and $dF( y(t) \mid \Delta N(t)=0, \overline{\boldsymbol{O}}(t) )$ be the distributions of $Y(t)$ among those participants who did, and who did not, have an assessment at $t$, given the observed past $\overline{\boldsymbol{O}}(t)$.  In each arm, let $\mu(t)=E[Y(t)]$, the mean of the (possibly unobserved) outcome at time $t$ among all individuals in the given arm.

\subsection{Observed data modeling and estimation}\label{sub:methods/modeling}

AIIW estimators require the user to fit two models to the observed data, separately in each arm:  an assessment-time intensity model, and an observed outcome distribution model.  Both models are separate from the sensitivity parameter (described in Section \ref{sub:marginalmeanmodel}) and therefore only need to be fit once in each treatment arm;  this is in keeping with the philosophy that all observed data modeling should be separate from the sensitivity parameters \citep{scharfstein1999nonignorable,franks2020observationalsa}.

\noindent \textbf{Assessment-time intensity model.}  As with previous IIW methods, AIIW estimators use weights based on the assessment-time intensity function given the observed past,
\[ \lambda( t \mid \overline{\boldsymbol{O}}(t)) = \lim_{\epsilon \to 0^+} \frac{P( N(t+\epsilon)-N(t) = 1 \mid \overline{\boldsymbol{O}}(t)) }{ \epsilon} . \]
The \CRANpkg{SensIAT} package uses a stratified Andersen-Gill model \citep{andersengill1982}, 
\[ \lambda( t \mid \overline{\boldsymbol{O}}(t))= \lambda_{0,k}(t) \exp\{ \boldsymbol{\gamma}^\prime \boldsymbol{Z}(t) \}  D_k(t),\] 
where $k$ denotes assessment number, $\lambda_{0,k}(t)$ is an unspecified baseline intensity function for stratum $k$, $\boldsymbol{\gamma}$ is a parameter vector,  and $D_k(t)$ is an indicator that the individual is at risk for having the $k$th assessment at time $t$; here $\boldsymbol{Z}(t)$ is a function of the observed past that is specified by the user, and may include factors such as outcomes at previous assessments and other key baseline or time-varying covariates.

The parameter vector $\boldsymbol{\gamma}$ is estimated using partial likelihood \citep{cox1972regression,cox1975partiallikelihood}, and the baseline intensity functions $\lambda_{0,k}( t )$ are estimated by kernel-smoothing the Breslow estimators \citep{breslow1972discussion, wells1994kernelestimation} of the cumulative baseline intensity functions.

\medskip

\noindent \textbf{Observed outcome distribution model.} Next, the conditional distribution of the observed outcomes given the observed past must be modeled.  The \CRANpkg{SensIAT} package uses a single index model \citep{CHIANG2012271}:
\[ F_{Y(t)} \big( y(t) \mid \Delta N(t)=1, \overline{\boldsymbol{O}}(t) \big) = G \big( y(t) \mid \boldsymbol{\theta}^\prime \boldsymbol{X}(t) \big), \]
where $G(\cdot|\cdot)$ is a valid conditional distribution function.
 The single index model is a flexible semiparameteric model that assumes that the conditional distribution function of an outcome, given a vector of predictors, depends on the predictors only through a univariate term determined by some vector $\boldsymbol{\theta}$. Here $\boldsymbol{X}(t)$ is a function of the observed past that is specified by the user, and may include factors such as outcome at the previous assessment, time of current assessment, and elapsed time since the previous assessment ("lag time").  Users should take care to avoid collinearity when selecting predictors, as this can result in problems when fitting the single index model.

 Let $Y_{ik}=Y_i(T_{ik})$, $\boldsymbol{X}_{ik}=\boldsymbol{X}_i(T_{ik})$. Given a kernel function $\kappa(\cdot)$, bandwidth $h$, and value of $\boldsymbol{\theta}$, define a cross-validated error term $e_{ik}( z; h, \boldsymbol{\theta}) = I(Y_{ik} \leq z) - \widehat F_{ik}(z ; h,\boldsymbol{\theta})$ where $I( \cdot )$ is the indicator function and $\widehat F_{ik}(z ; h,\boldsymbol{\theta})$ is the Nadaraya-Watson estimator
\begin{equation}\label{eqn:CV_hatF}
    \widehat F_{ik}(z ; h,\boldsymbol{\theta}) = \frac{
        \sum_{j=1}^n \sum_{\ell = 1}^{K_j}
        I(Y_{j \ell} \leq z)
        I(j\neq i)
        \kappa(( \boldsymbol{X}_{j \ell} - \boldsymbol{X}_{ik})^\prime \boldsymbol{\theta}/h) 
    }{
        \sum_{j=1}^n \sum_{\ell = 1}^{K_j}
        I(j\neq i)
        \kappa(( \boldsymbol{X}_{j \ell} - \boldsymbol{X}_{ik})^\prime \boldsymbol{\theta}/h)
    }.
\end{equation}
We define the pseudo sum of integrated squared error (PSIS) as $$PSIS(\boldsymbol\theta,h) = \sum_{i=1}^n \sum_{k = 1}^{K_i}   \sum_{j=1}^n \sum_{\ell = 1}^{K_j} e_{ik}^2( Y_{j \ell}; h,\boldsymbol{\theta}).$$  The software provides three options for using $PSIS(\boldsymbol\theta,h)$ to estimate $\boldsymbol\theta$ and $h$; see Section \ref{ObservedOutcomesModel}. We denote the estimators for $\boldsymbol\theta$ and $h$ as $\widehat{\boldsymbol\theta}$ and $\widehat{h}$, respectively.

The AIIW estimator incorporates estimates of $F\big( y(t) \mid \Delta N(t)=1, \overline{\boldsymbol{O}}_i(t) \big)$ for each individual $i$ for each time $t$, where $\overline{\boldsymbol{O}}_i(t)$ is this individual's observed past prior to time $t$.  This is estimated using the Nadaraya-Watson estimator with the estimated $\widehat{\boldsymbol{\theta}}$ and $\widehat{h}$, 

\[ \widehat F( y(t) \mid \Delta N(t)=1, \boldsymbol{X}_i(t) ) = 
    \frac{
        \sum_{j=1}^n \sum_{\ell = 1}^{K_j}
        I(Y_{jl} \leq y(t)) \kappa( ( \boldsymbol{X}_{j \ell } - \boldsymbol{X}_i(t))^\prime \widehat{\boldsymbol{\theta}}/\widehat{h}) 
    }{
        \sum_{j=1}^n \sum_{\ell = 1}^{K_j}
        \kappa(( \boldsymbol{X}_{j \ell} - \boldsymbol{X}_i(t))^\prime \widehat{\boldsymbol{\theta}}/\widehat{h})
    }. \]

\subsection{AIIW estimation of the marginal mean}
\label{sub:marginalmeanmodel}

\textbf{Marginal mean model.}  The method of \citet{smith2024sa} assumes a population marginal mean model, separately for each treatment arm. The user selects a time interval or intervals for the marginal mean model.  If, in their study, assessments take place continuously throughout some interval that contains all target times $t^*$, the user may specify a single interval $[t_1,t_2]$.  In this case, the marginal mean model is
\begin{equation} \label{marg_mean} 
\mu(t) = \boldsymbol{B}(t)^\prime \boldsymbol{\beta} \mbox{  for  } t \in [t_1,t_2], 
\end{equation}
where $\boldsymbol{B}(t)$ is a specified vector-valued function of time and $\boldsymbol{\beta}$ a parameter vector.  The \CRANpkg{SensIAT} package takes $\boldsymbol{B}(t)$ to be a spline basis.  Knots are specified by the user;  the first and last knots should be placed at the beginning and end of the interval. 

Alternatively, in some studies, assessments may be concentrated in two or more disjoint intervals, separated by substantial gaps in time during which few assessments occur.  In this case the user should specify multiple intervals, say $[t_{m1},t_{m2}]$ for $m=1,\ldots, M$, so that inference is drawn only for time periods that are well-supported by the study's data.  In this case, the marginal mean model is
\[ \mu(t)=\boldsymbol{B}_m(t)^\prime \boldsymbol{\beta}_m \mbox{ for } t \in [t_{m1},t_{m2}], \mbox{ for } m=1,\ldots, M, \]
where $\boldsymbol{B}_m(t)$ and $\boldsymbol{\beta}_m$ are interval-specific versions of $\boldsymbol{B}(t)$ and $\boldsymbol{\beta}$, respectively.  The parameters $\boldsymbol{\beta}_m$ are estimated separately for each interval.  In this case, a list of knot sequences is specified by the user;  the first and last knots of the $m$th sequence should be placed at the beginning and end of the $m$th time interval.

\medskip

\noindent \textbf{Sensitivity parameter.}  In the model above, $\mu(t)$ is the population mean outcome in the given arm at time $t$, averaged over all individuals who were not assessed at time $t$ as well as those who were.  Inference is made under an assumption that connects $dF( y(t) \mid \Delta N(t)=0, \overline{\boldsymbol{O}}(t))$ and $dF( y(t) \mid \Delta N(t)=1, \overline{\boldsymbol{O}}(t))$, using a sensitivity parameter $\alpha$ for each arm.  A value of $\alpha >0$ indicates that unobserved outcomes tend to be higher than observed outcomes within each stratum of $\overline{\boldsymbol{O}}(t)$, while a value of $\alpha <0$ indicates that unobserved outcomes tend to be lower than observed outcomes, and $\alpha =0$ assumes that there is no difference between the distribution of unobserved versus observed outcomes.  See \citep{smith2024sa}, Figure 2, for an illustration.  IIW methods \citep{lin2004dependent,buzkova2007dependent,buzkova2009repeated,pullenayegum2013dr,sun2016quantile} deal with the special case of $\alpha=0$, which has been referred to as explainable assessment \citep{smith2024sa}, assessment at random \citep{pullenayegum2022repeatedly}, or visiting at random \citep{pullenayegum2016review}. 

\medskip

\noindent \textbf{AIIW estimation.}  Augmented inverse intensity weighted estimators use inverse weighting by the intensity function

\[ \rho( t \mid Y(t), \overline{\boldsymbol{O}}(t)) = \lim_{\epsilon \to 0^+} \frac{P( N(t+\epsilon)-N(t) = 1 \mid Y(t), \overline{\boldsymbol{O}}(t)) }{ \epsilon}, \]
reflecting the fact that, in our setting, assessment can be impacted by the current outcome value $Y(t)$ as well as the observed past.  \citet{smith2024sa} have shown that, for each value of $\alpha$, $\rho( t \mid Y(t), \overline{\boldsymbol{O}}(t) ; \alpha)$ is determined by the intensity $\lambda( t \mid  \overline{\boldsymbol{O}}(t))$ and observed outcome distribution $dF \big( y(t) \mid \Delta N(t)=1, \overline{\boldsymbol{O}}(t) \big)$ modeled in the previous section:  
\[  \rho(t \mid Y(t), \overline{\boldsymbol{O}}(t); \alpha) =  
    \lambda(t \mid \overline{\boldsymbol{O}}(t)) \frac{ E \big[ \exp \{ \alpha Y(t) \} \mid \Delta N(t)=1,\overline{\boldsymbol{O}}(t) \big] }{ \exp\{ \alpha Y(t) \} }. \]
They have additionally shown that, for each $\alpha$, $E[ Y(t) \mid \boldsymbol{O}(t)]$ is determined by the observed outcome distribution:  
\[ E[ Y(t) \mid \overline{\boldsymbol{O}}(t); \alpha ] = 
    \frac{ E[ Y(t) \exp\{ \alpha Y(t) \}\mid \Delta N(t)=1, \overline{\boldsymbol{O}}(t) ] }{ E[ \exp\{\alpha Y(t)\}\mid \Delta N(t)=1, \overline{\boldsymbol{O}}(t) ] }. \]

In the case where a marginal mean model is specified on a single interval $[t_1,t_2]$, the AIIW estimator of $\boldsymbol{\beta}$ under a given value of $\alpha$, based on data from $n$ independent individuals in the given treatment arm, is
\[ \widehat{\boldsymbol{\beta}} = \frac{1}{n} \sum_{i=1}^n  \left\{ \sum_{k=1}^{K_i} \boldsymbol{V}^{-1} \boldsymbol{B}(T_{ik} )\frac{ \big( Y_i(T_{ik}) - \widehat{E}[ Y(T_{ik}) \mid \overline{\boldsymbol{O}}_i(T_{ik}); \alpha ] \big)}{ \widehat{\rho}( T_{ik} \mid Y_i(T_{ik}),\overline{\boldsymbol{O}}_i(T_{ik}); \alpha)} + \int_{t=t_1}^{t_2}  \boldsymbol{V}^{-1} \boldsymbol{B}(t) \widehat{E}[ Y(t) \mid \overline{\boldsymbol{O}}_i(t); \alpha ] dt \right\}\]
where
\begin{eqnarray}\label{eqn:expectedvalues}
  \widehat{E}[ Y(t) \mid \overline{\boldsymbol{O}}_i(t); \alpha ] &=& 
    \frac{
        \widehat{E}[ Y(t) \exp \{\alpha Y(t) \}\mid \Delta N(t)=1,\overline{\boldsymbol{O}}_i(t) ]
    }{
        \widehat{E}[ \exp \{\alpha Y(t)\}\mid \Delta N(t)=1,\overline{\boldsymbol{O}}_i(t) ]
    },\\ 
  \widehat\rho(t \mid Y_i(t), \overline{\boldsymbol{O}}_i(t); \alpha) &=&  
    \widehat\lambda(t \mid \overline{\boldsymbol{O}}_i(t)) \frac{ \widehat{E} \big[ \exp \{ \alpha Y(t) \} \mid \Delta N(t)=1,\overline{\boldsymbol{O}}_i(t) \big] }{ \exp\{ \alpha Y_i(t) \} },
\end{eqnarray}
and $\boldsymbol{V}=\int_{t_1}^{t_2} \boldsymbol{B}(t) \boldsymbol{B}(t)^\prime dt$.  In the case where multiple disjoint intervals are used for the marginal mean model, the parameter for each interval is estimated separately.  For the $m$th interval, $[t_{m1},t_{m2}]$,  
\begin{align*}   
\widehat{\boldsymbol{\beta}}_m = 
& \frac{1}{n} \sum_{i=1}^n  \left\{ 
    \sum_{k=1}^{K_i} I( T_{ik} \in [t_{m1},t_{m2}]) \boldsymbol{V}_m^{-1} \boldsymbol{B}_m(T_{ik} )\frac{ \big( Y_i(T_{ik}) - \widehat{E}[ Y(T_{ik}) \mid \overline{\boldsymbol{O}}_i(T_{ik}); \alpha ] \big)}{ \widehat{\rho}( T_{ik} \mid Y_i(T_{ik}),\overline{\boldsymbol{O}}_i(T_{ik}); \alpha)} + 
\right. \\
 & \left. 
    \phantom{\frac{1}{n} \sum_{i=1}^n  \left\{\right.}
    \int_{t=t_{m1}}^{t_{m2}}  \boldsymbol{V}_m^{-1} \boldsymbol{B}_m(t) \widehat{E}[ Y(t) \mid \overline{\boldsymbol{O}}_i(t); \alpha ] dt \right\}, \end{align*}
where $\boldsymbol{V}_m=\int_{t_{m1}}^{t_{m2}} \boldsymbol{B}_m(t) \boldsymbol{B}_m(t)^\prime dt$.  For notation let
\[
\widehat{\boldsymbol{\beta}}_m = \frac{1}{n} \boldsymbol{V}_m^{-1}\sum_{i=1}^n \left\{
    \phi_{m,i,1}(\alpha) + \phi_{m,i,2}(\alpha)
\right\}
\]
with 
\begin{eqnarray}
\label{eqn:influence1}
\phi_{m,i,1}(\alpha) &=& \sum_{k=1}^{K_i}
    I( T_{ik} \in [t_{m1},t_{m2}])
    \boldsymbol{B}_m(T_{ik} )\frac{ 
        \big( Y_i(T_{ik}) - \widehat{E}[ Y(T_{ik}) \mid  \overline{\boldsymbol{O}}_i(T_{ik});\alpha ] \big)
    }{ 
        \widehat{\rho}( T_{ik} \mid  Y_i(T_{ik}),\overline{\boldsymbol{O}}_i(T_{ik}); \alpha)
    }\mathrm{, and} \\
\label{eqn:influence2}
\phi_{m,i,2}(\alpha) &=& \int_{t=t_{m1}}^{t_{m2}}  
    \boldsymbol{B}_m(t) 
    \widehat{E}[ Y(t) \mid  \overline{\boldsymbol{O}}_i(t);\alpha ]
    dt
\end{eqnarray}
corresponding to the inverse-weighted term and the augmentation term, respectively.  We refer to these as the influence function terms.

\subsection{Variance estimation and confidence intervals}

\citet{smith2024sa} have shown asymptotic normality of $\widehat{\boldsymbol{\beta}}$.  The \CRANpkg{SensIAT} package implements jackknife variance estimation for the variance of $\widehat{\boldsymbol{\beta}}$ and $\widehat{\mu}(t) = \boldsymbol{B}(t)^\prime \widehat{\boldsymbol{\beta}}$, to be used in constructing Wald confidence intervals for $\mu(t)$. Simulation studies have found that such confidence intervals perform better in terms of coverage than Wald confidence intervals using IF-based standard errors.

\subsection{Sensitivity analysis: mean outcome in each arm}

To conduct the sensitivity analysis, \citet{smith2024sa} suggest eliciting values $\mu_{min}$ and $\mu_{max}$ from a domain expert that would be implausibly low/high for the mean outcome at any time.  For each treatment arm, the analyst should obtain mean outcome curves, $\widehat{\mu}(t) = \boldsymbol{B}(t)^\prime \widehat{\boldsymbol{\beta}}$ for $t \in [t_1,t_2]$, under a broad range of $\alpha$ values.  The analyst should then restrict the ranges of $\alpha$ for that arm to include only those values that correspond to mean curves that lie completely between $\mu_{min}$ and $\mu_{max}$.

\subsection{Sensitivity analysis: treatment effect}

To compare treatment arms, let $a=1$ denote treatment and $a=0$ denote control, and we use superscript-$(a)$ to denote the given arm.  For the treatment effect, we focus on $\delta(t^*)=\mu^{(1)}(t^*)-\mu^{(0)}(t^*)$ at each target time $t^*$.  To conduct the sensitivity analysis for $\delta(t^*)$, the analyst should estimate $\delta(t^*)$ under a grid of $\alpha^{(1)}$ and $\alpha^{(0)}$ values in the restricted ranges determined above.

\section{Implementation}

\subsection{Overview} \label{subsec:overview}
As explained in Section \ref{sub:methods/modeling}, in \CRANpkg{SensIAT} two models must be specified:  one for the assessment times, referred to as the intensity model, and one for the conditional distribution of the observed outcomes given the observed past, referred to as the observed outcomes model. Models are assumed to have the same structure across treatment arms, but are fit separately to each arm.

The package \CRANpkg{SensIAT} was built in R with critical performance sections written in C++.
The topmost function that fits the models for each arm of the data is \code{fit\_SensIAT\_fulldata\_}\-\code{model()}.  This function splits the data into treatment and control groups and passes the fit to \code{fit\_SensIAT\_within\_group}\-\code{\_model()}, which is the workhorse function of the package.  This fits the intensity model, the observed outcomes model, and the marginal mean model given the sensitivity parameter, $\alpha$.

Because the intensity and observed outcomes models are linked with common variables, the following variables are specified individually:
\begin{itemize}
    \item \code{id}, the patient/subject identifier,
    \item \code{outcome}, the outcome variable, and 
    \item \code{time}, the time of observations.
\end{itemize}
These variables are internally transformed into variables used in the models:
\begin{itemize}
    \item \code{..time..} and \code{..outcome..} are just renames of the specified variables,
    \item \code{..visit\_number..}, inferred from the \code{time} variable within \code{id} only used for stratification in the intensity model,
    \item \code{..prev\_outcome..}, the previous observed outcome,
    \item \code{..prev\_time..}, the time of the previous assessment,
    \item and \code{..delta\_time..} the time since the previous assessment ("lag time").
\end{itemize}
These are used to construct the model formulae and carry special meaning in the context of the marginal mean model. 

The arguments \code{intensity.args}, \code{outcome.args}, and \code{influence.args} of the function \\ \code{fit\_SensIAT\_within\_group\_model()} control the fitting of the intensity model, the observed outcomes model, and the marginal mean model, respectively. Each should be a list.  Details of the contents of these arguments will be given in their respective sections. 

\subsection{Data formatting} \label{subsec:formatting}

Users should structure their data as a data frame in long format with a row for each assessment, with at least columns for participant ID; treatment arm;  time of the assessment, \texttt{time}; and outcome at that time, \texttt{outcome}.  See Table \ref{tab_data} for an example.  Users may also include additional columns for any baseline covariates and/or time-varying covariates that they wish to include in the intensity model and/or the observed outcomes model.  However, any time-varying covariates must be variables whose values were known at the time of the previous assessment.  For example, users should not include a column \code{BP} that contains a value for blood pressure at the current assessment, \texttt{time};  but they may create a column \code{Previous\_BP} that contains a value for blood pressure recorded at the previous assessment.

There are two options for how users can specify when participants remain on-study.  If using the default (\code{add.terminal.observations = TRUE}), then each row in the user-supplied data frame should correspond to an assessment and there cannot be any NA values of \texttt{outcome}.  The package internally adds ``non-event" rows to indicate when participants are still at risk for subsequent assessments.  This default option makes two assumptions about the design and course of the study:

\begin{itemize} 
\item There was a fixed maximal number of follow-up visits that participants could have, and this was realized by at least one participant.  
\item All participants who had fewer than the maximal number of visits remained on-study for the same length of time, specified as the \texttt{End} parameter. In particular, these individuals remained at risk for a further assessment until the time \texttt{End}, whereas individuals who had the maximal number of visits left the risk set at the time of their last visit.

\end{itemize}

 \noindent Additionally, this default option should only be used if \texttt{..prev\_outcome..} is to be included in the intensity model (as in the default model formula described in Section \ref{subsec:intensity}).
 
 Alternatively, under the second option, all information about remaining on-study is specified explicitly by the user. This allows flexibility for a study where, for example, there was no maximal number of visits in the study design (i.e. where every participant remained on-study after their last assessment), or where participants who were recruited later in calendar time had a shorter length of follow-up than participants who were recruited earlier in calendar time.  With this option, the user-supplied data frame should include a terminal row for each participant who remained on-study after their last assessment.  In this terminal row, \texttt{time} should be the time at which the individual left the study and \texttt{outcome} needs to be set to NA;  the NA value of \texttt{outcome} serves as the indicator that there was no event at this time.  The option for the package to internally add "non-event" rows should be turned off with \code{add.terminal.observations = FALSE}.

\subsection{Intensity model} \label{subsec:intensity}

The intensity model for assessment times is an Andersen-Gill model. By default, the only predictor variable is the most recent observed outcome and the model is stratified by visit number, i.e. the number of observations that have occurred. The predictor variables are modifiable by the user through the \code{"model.modifications"} element of the \code{intensity.args} list.  This should be a formula compatible with the \code{update.formula()} function, which can either update or overwrite the default model formula, 

\begin{verbatim}Surv(..prev_time.., ..time..,  !is.na(..outcome..)) ~ ..prev_outcome.. + 
                                            strata(..visit_number..)
\end{verbatim}
For example, to add the standard demographics variables \code{Age} and \code{RaceEthnicity} to the intensity model specification, the following would be included in the function call:
\begin{verbatim}
intensity.args = list(
    model.modifications = . ~ . + Age + RaceEthnicity
)
\end{verbatim}  
See the caution in Section \ref{subsec:formatting} on time-varying covariates:   only covariates whose values were known at \texttt{..prev\_time..} can be included in the intensity model.  Once the Andersen-Gill model 
is fit, the intensity is estimated by computing the discrete hazards then smoothing over time with a user provided kernel (\code{intensity.args\$kernel}) and bandwidth (\code{intensity.args\$bandwidth}). By default, the Epanechnikov kernel is used, and bandwidth estimated via the \code{dpill()} function from the \code{KernSmooth} package.

\subsection{Observed outcomes model} \label{ObservedOutcomesModel}

The current implementation of the package supports a single index model for the conditional distribution of the observed outcomes. The default formula is 
\begin{verbatim}
..outcome.. ~ -1 + ns(..prev_outcome.., df=3) +
             scale(..time..) + scale(..delta_time..)
\end{verbatim}
The form of the observed outcomes model is controlled by passing in a function for fitting the model in the \code{outcome\_modeler} argument. 
 The \code{outcome\_modeler} must accept a formula for the first argument and a \code{data} argument by name. Similar to the intensity model, the \code{outcome.args} may include a \code{model.modifications} element with a formula to modify the default formula prior to it being passed into the \code{outcome\_modeler}. An example of this is shown in Section \ref{sec:HAP2}. All elements of \code{outcome.args} except \code{model.modifications} will be passed to the \code{outcome\_modeler()} function as additional arguments.  Predictors that may be included in the observed outcomes model include baseline covariates, outcome values observed at previous assessment times such as \texttt{..prev\_outcome..}, and any functions of these.  Time-varying covariates or functions of these can be included, as long as their (recorded) values change only at assessment times and their values were known at \texttt{..prev\_time..}, as noted in Section \ref{subsec:formatting}.  The variables \texttt{..time..} and \texttt{..delta\_time..}, or scaled versions of these, can be included.  However, for reasons of computational efficiency (see Section \ref{subsec:marginal}), the only functions of \texttt{time} that the \CRANpkg{SensIAT} package allows as predictors in the observed outcomes model are piecewise linear functions with a common slope on each piece, that may have jump discontinuities at assessment times, but that are continuous elsewhere.  In particular, terms such as \texttt{log(time)}, \texttt{ns(time)}, or \texttt{time}$^2$ cannot be used.

The single index outcome model laid out in Section \ref{sub:methods/modeling} is fit by minimizing $PSIS(\boldsymbol{\theta},h)$ with respect to $\boldsymbol{\theta}$ and $h$, however this minimization is over-parameterized by one degree of freedom.
We considered three constraints to address this issue: (1) setting the first coefficient of $\boldsymbol{\theta}$ to 1 (the original restriction used in \citet{smith2024sa}), (2) setting the bandwidth $h$ to 1, and (3) restricting the norm of $\boldsymbol{\theta}$ to 1. We have implemented three corresponding methods. In each of these methods, by default, we use an estimator of $\boldsymbol{\theta}$ using the Minimum Average Variance Estimation (MAVE) method \citep{xia2002adaptive,wang2008sliced} as the initial value. We found that some data sets were unable to converge with one parameterization while others succeeded, so all three options are available. 

The first method initializes $\boldsymbol{\theta}$ at a scaled version of the MAVE estimator, scaled to have first element one, and initializes the bandwidth $h$ at 1, then minimizes $PSIS(\boldsymbol{\theta},h)$ with respect to $\boldsymbol{\theta}$ and $h$. This option is implemented in the function \code{fit\_SensIAT\_single\_index\_fixed\_coef\_model()}. The second proposed method initializes $\boldsymbol{\theta}$ at the MAVE estimator, holds the bandwidth $h$ fixed at 1, and minimizes $PSIS(\boldsymbol{\theta},1)$ with respect to $\boldsymbol{\theta}$. This option is implemented in the function \code{fit\_SensIAT\_single\_index\_fixed\_bandwidth\_model()}. The last option minimizes $PSIS(\boldsymbol\theta,h)$ with respect to $\boldsymbol{\theta}$ and $h$, under the restriction that $\boldsymbol{\theta}$ has norm 1. Each step of the optimization procedure alternates between minimizing with respect to $h$ and  minimizing with respect to $\boldsymbol{\theta}$, under the norm 1 restriction.  When minimizing with respect to $h$, the standard functions \code{stats::optimize()} or \code{stats::optim(method="L-BFGS-B")} (the default) or a grid search may be used;  this is specified via the \code{bw.method} option.  In order to obtain an empirical rule for reasonable bounds for $h$, we parametrize the bandwidth as a multiple of the standard deviation $
\widehat{\sigma}=\sqrt{\mathrm{var}(X^\prime \widehat{\boldsymbol{\theta}})}$, as $h=h^\star \widehat{\sigma}$, and we take $h^\star\in[0.01,1.5]$;  this is modifiable via the \code{bw.range} option. After this initial estimate of $\widehat{\boldsymbol{\theta}}$, further iterations can be performed using the \code{manifold.optim()} function from the \CRANpkg{ManifoldOptim} package \citep{ManifoldOptim}. 
 This parameterization gives a geometric interpretation of the model where $\boldsymbol{\theta}$ specifies a direction, or relative weighting, and $h^\star$ specifies the magnitude or width of the kernel. This option is implemented in the function \code{fit\_SensIAT\_single\_index\_norm1coef\_model()}.

All three options for fitting the single index model return list objects with at least the elements \code{coefficients}, \code{bandwidth}, and \code{details}, which gives the details from the optimization routine such as criteria reached.  The returned objects have an appropriate \code{S3} class added, as well as recording the subject identifier variable and the kernel used. For this outcome model class we implemented the standard model generics \code{coef()}, \code{formula()}, \code{model.frame()}, \code{model.matrix()} and \code{predict()}. We also took advantage of the \code{generics} package to implement \code{generics::prune()}, which strips out unneeded elements to conserve space when computing the jackknife variance estimations.

\subsection{Marginal mean model and influence function computations} \label{subsec:marginal}

The final step is to fit the marginal mean model with the information given from the intensity model and the observed outcomes model. We recommend that users fit the model through the \code{fit\_SensIAT\_fulldata\_mode7l()} or \code{fit\_SensIAT\_within\_group\_model()} functions, because of the tight coupling between the variables in the models. Internally, \CRANpkg{SensIAT} computes the influence function terms $\phi_{m,i,1}(\alpha)$ and $\phi_{m,i,2}(\alpha)$ defined in Equations \eqref{eqn:influence1} and \eqref{eqn:influence2}. The function \code{compute\_influence\_}\-\code{terms()} is a generic dispatching on the outcome model so that new models that are in development may be easily added later. It expects the observed outcomes model and the intensity model; the sensitivity parameter, $\alpha$; the data used to fit the intensity model and the observed outcomes model; and the spline basis object, from package \code{orthogonalsplinebasis} \citep{redd2012comment}.  It is expected to return a \code{data.frame} object with one row per patient with the columns \code{id}, \code{term1}, and \code{term2}, where the latter two columns are complex---either lists or matrices---since each term should have multiple columns. These terms are then used to compute $\widehat{\boldsymbol{\beta}}$ of the marginal mean model.

The computation of $\phi_{m,i,1}$ is straightforward, with estimates of the expected values \\ $E[ Y(T_{ik}) \exp\{\alpha Y(T_{ik})\} \mid \Delta N(T_{ik})=1,\overline{\boldsymbol{O}}_i(T_{ik}) ]$ and $E[\exp\{ \alpha Y(T_{ik})\} \mid \Delta N(T_{ik})=1,\overline{\boldsymbol{O}}_i(T_{ik}) ]$ computed using the estimate of $F\big( y(T_{ik}) \mid \Delta N(T_{ik})=1, \overline{\boldsymbol{O}}_i(T_{ik}) \big)$ at each of the participant's observation times $T_{ik}$. The estimate of $\lambda(T_{ik} \mid \overline{\boldsymbol{O}}_i(T_{ik}))$ 
is computed as previously outlined.

Computation for $\phi_{m,i,2}$ is more complicated, as it involves integration of $\boldsymbol{B}_m(t)\widehat{E}[ Y(t) \mid \overline{\boldsymbol{O}}_i(t); \alpha ]$ over the entire interval of interest. For each time in this interval, the estimated conditional mean outcome at time $t$ given the participant's observed past before $t$ may involve time itself and possibly functions of time, such as time since last observation.  The assumptions imposed in Section \ref{ObservedOutcomesModel} on the predictors used in the observed outcomes model---i.e.\ piecewise linearity in time with possible discontinuities only at assessment times---are leveraged here for speed, and 
the aforementioned variables \code{..time..} and \code{..delta\_time..} have special meaning in the computations here.  Practically, the integration is split into distinct intervals separated by assessment times, as the integrand is smooth inside each of these intervals and discontinuous at the assessment times.  
Although the predictors are linear in time, the integrand in $\phi_{m,i,2}$ is not, preventing simplification or algebraic solutions, which necessitates utilizing numerical integration.

To perform the numerical integration, one of two methods is used.  The first is a fixed-width trapezoidal approximation (\code{influence.args\$method='fixed'}), as was originally presented in \citet{smith2024sa}.  The second method, which is the default, is a vectorized adaptive Simpson quadrature (\code{influence.args\$method='adaptive'}), adapted from the \CRANpkg{pracma} R package \citep{pracma} for our purpose and compiled for speed.  For the trapezoidal numerical integration, you may specify either the width (\code{influence.args\$delta}) or the number of points to specify (\code{influence.args\$resolution}).  The adaptive integration criteria is controlled with \code{influence.args\$tolerance}.

\subsection{Wrap-up}

The \code{fit\_SensIAT\_within\_group\_model()} function returns the fitted models as a \code{SensIAT\_within\_}\-\code{group\_model} object.  The 
\code{fit\_SensIAT\_fulldata\_model()} function returns a list with elements for \code{control} and \code{treatment} each of which are \code{SensIAT\_within\_group\_model} objects and given the class \code{SensIAT\_fulldata\_model}. 

For these top-level objects there are three primary functions. The first is the standard \code{predict()} function, which here additionally requires a time.  For a \code{SensIAT\_within\_group\_model} object, a \code{data.frame} is returned with the sensitivity parameter $\alpha$, \code{alpha}; the time, \code{time}; the estimated marginal mean, $\widehat{\mu}(t)$, \code{mean}; and the IF-based estimate of the variance of $\widehat{\mu}(t)$, \code{var}. For a \code{SensIAT\_fulldata\_model} object, the returned \code{data.frame} has the time; $\alpha$, $\widehat{\mu}(t)$, and the IF-based estimate of the variance of $\widehat{\mu}(t)$ for each arm;   the estimated treatment effect, $\widehat{\delta}(t)$, \code{mean\_effect}; and the IF-based estimate of the variance of $\widehat{\delta}(t)$,  \code{var\_effect}.

Previously \citet{smith2024sa} found that the IF-based variance estimator resulted in poor confidence interval coverage compared to the jackknife variance estimator. We include a generic function \code{jackknife()} which can compute the jackknife variance for the within group object or for the full data model, which computes the jackknife on each arm separately. Either returns a \code{data.frame} with the mean and variance estimated through the jackknife, along with the other values computed with \code{predict()}. The \code{jackknife()} function for the full model also includes the \code{mean\_effect\_jackknife\_var} for the jackknife estimated variance of the treatment effect.

The resulting tables are appropriately given a class to facilitate passing to the last function, \code{ggplot2::autoplot()}. The four methods provided give four different plots.  For the within group object, a line plot for the estimated marginal mean at each time is produced, with alpha differentiated by color. For the jackknife, a dot and whisker plot is produced, with a dot at the point estimate and error bars showing a 95\% Wald confidence interval, using the jackknife variance estimate. Again, alpha is differentiated by color. A \code{ggplot2::position\_dodge} is included by default, as it is expected that the dot and whiskers will overlap, however the width can be specified.  The plots for the full model are both color contour plots, with the horizontal axis representing alpha for the control and the vertical for the treatment. For the full model, the color indicates the estimated treatment effect at the given pair of sensitivity values.  For the full model jackknife results, the color represents either, zero if the resulting 95\% confidence interval contains zero or the bound of the 95\% confidence interval closest to zero.  Both full model plots are for a single point in time; however, if multiple time points are included, facet wrapping is included.  All four methods produce standard \code{ggplot} objects to which further customizations are handled in the typical \code{ggplot2} manner.

\section{Data analysis}
\label{sec:HAP2}

\subsection{Synthetic HAP2 data}
The Helping Asthma Patients 2 (HAP2) study \citep{apter2020patientadvocates} aimed to evaluate whether a patient advocate (PA) intervention could improve asthma outcomes compared to usual care (UC) in adults with moderate to severe asthma.  Patients were recruited from clinics serving low-income urban neighborhoods.
A total of 312 eligible participants were randomized to receive either six months of the PA intervention ($n=156$) or UC ($n=156$). Both groups continued to receive asthma care from clinicians and practices that generally adhered to asthma management guidelines. Participants assigned to UC did not interact with PAs, nor did PAs accompany them to provider visits. In the intervention group, PAs—recent college graduates interested in health care careers—coached, supported, and helped participants prepare for asthma-related medical visits. They attended visits alongside participants and confirmed participants' understanding of provider recommendations.  All participants were scheduled to have asthma outcomes assessed at 12, 24, 36 and 48 weeks post-randomization.  Some participants agreed to also provide outcome data at 72 and 96 weeks. However, in practice there was substantial variability in the actual assessment times:  times of participants' 1st through 6th post-baseline assessments had mean (standard deviation) 18.2 (13.5), 34.1 (14.1), 48.1 (15.9), 64.0 (17.8), 128.2 (25.6), and 155.5 (27.7) weeks in the PA arm, and 17.2 (9.2), 36.0 (19.0), 49.4 (14.1), 64.1 (16.6), 125.7 (25.7), and 152.7 (26.6) weeks in the UC arm.  Therefore, bias due to possible informative timing could be a concern, and a sensitivity analysis would be valuable.

\begin{figure}
\begin{center}

\includegraphics[width=0.8\textwidth]{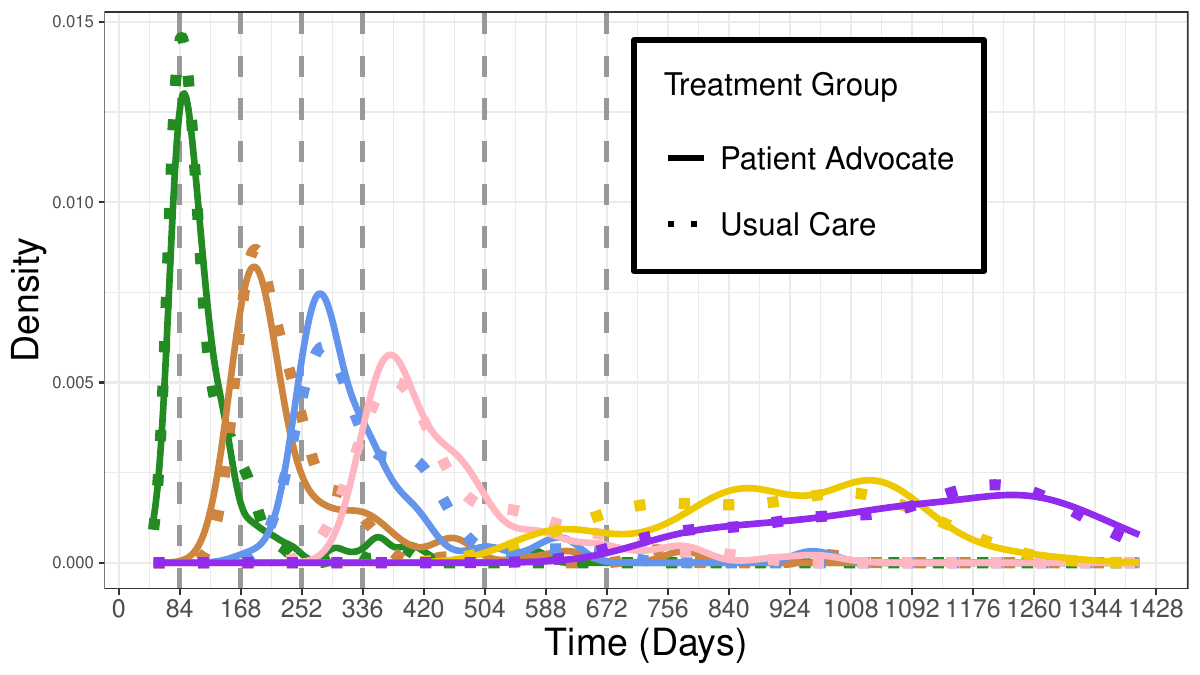}

\end{center}

\caption{Times of post-baseline assessments in each arm in the synthetic HAP2 data.  Assessments are color-coded by visit number.}

\label{fig:times}

\end{figure}

We constructed a synthetic dataset based on the HAP2 data, using the procedure described in Section \ref{sec:sim_study}, with sample size $n=156$ for each treatment arm.   For this analysis, we focused on asthma control, measured by the 6-item Asthma Control Questionnaire (ACQ), which reflects symptoms over the past week. Each item on the ACQ is scored from 0 (completely controlled) to 6 (extremely uncontrolled). Assessment times (in days), stratified by treatment arm and visit number, are shown in Figure \ref{fig:times}; summary information is given in Table \ref{tbl:times}.  

We now illustrate use of the software to analyze the HAP2 synthetic data.  Specifically, we seek to compare treatment groups with respect to the marginal mean ACQ scores at 6 and 12 months. We use the superscript $(a)$ to denote treatment group, where $a=0$ denotes UC and $a=1$ denotes PA.

\subsection{Analysis of the synthetic HAP2 data}

\begin{table}

\small

\begin{center}

\begin{tabular}{c|ccc|ccc}
Visit & \multicolumn{3}{c|}{\textbf{Patient Advocate}} & \multicolumn{3}{c}{\textbf{Usual Care}} \\
number ($k$) &  $n$ & Mean Time (days) & Mean ACQ & $n$ & Mean Time (days) & Mean ACQ \\
 \hline
0 &  156 & 0 & 2.3 
& 156 & 0 & 1.9 \\
1 &  148 & 132 & 1.9 
& 154 & 116 & 1.7 \\
2 &  142 & 252 & 1.7 
& 147 & 251 & 1.7 \\
3 &  133 & 340 & 1.6 
& 136 & 346 & 1.7 \\
4 &  123 & 457 & 1.6 
& 119 & 455 & 1.6 \\
5 &  59 & 905 & 1.7 
& 69 & 883 & 1.7 \\
6 &  52 & 1101 & 1.7 
& 59 & 1082 & 1.7 \\ \hline
\end{tabular}
\end{center}

\normalsize

\caption{Summary information, stratified by visit number and treatment arm, for the synthetic HAP2 data.  The number of participants ($n$), the mean of the visit times, and the mean ACQ Score are shown.}

\label{tbl:times}

\end{table}

To define the time interval for estimation of the treatment-specific marginal mean curves, we selected the 2.5\% and 97.5\% quantiles of all post-baseline assessment times across both the UC and PA groups in the synthetic data, resulting in a range of $[76, 1232] $ days. For each group, we assumed:
\begin{align*}
    \mu^{(a)}(t) = \boldsymbol{B}(t)^\prime \boldsymbol{\beta}^{(a)}, \ 76\leq t \leq 1232, \ a = 0 \mbox{ (UC)},1 \mbox{ (PA)},
\end{align*}
where $t$ denotes day and $\boldsymbol{B}(t)$ is the basis of cubic B-splines with one interior knot at $t=654$ days.

We modeled the treatment-specific intensity functions  using a stratified Andersen-Gill model  as
$$\lambda^{(a)}\{t \mid \overline{\boldsymbol{O}}(t)\} = \lambda^{(a)}_{0,k}(t)\exp\{\gamma^{(a)} Z(t)\}D_k(t),\  k = 1,\ldots, 6, \ a=0,1,$$
where the stratification variable $k$ was the number of previous assessments and 
 $Z(t)$ was chosen to be the outcome at the previous assessment. We used kernel smoothing with an Epanechnikov kernel and a bandwidth of 30 days to estimate the treatment-specific baseline intensity functions. For the observed outcome distribution model, we used the single index model and chose the outcome at the previous assessment, current time and time since last observation as predictors.

Here we initially considered a range of $-0.7\leq \alpha^{(a)} \leq 0.7$ for the sensitivity parameter in each treatment group.

We applied the \code{fit\_SensIAT\_fulldata\_model()} function to estimate $\boldsymbol{\beta}^{(a)}$ for $a=0,1$. Table \ref{tab_data} displays the observations for the first participant in each treatment arm. Note that these participants have fewer assessment times than the maximum number of post-baseline visits (6). In this dataset, \code{ID} is the patient identifier, \code{outcome} is the ACQ score, \code{time} is the assessment time, and \code{Trt} indicates the treatment group.

\begin{table}[H]
\small
\centering
\begin{tabular}{c |c | c |c }
\hline
ID & outcome & time & Trt \\
 \hline 
 \hline
 1 & 3.3333333 &   0 & UC  \\
 1 & 2.5000000 &  76 &  UC \\
 1 & 2.5000000 & 162 & UC \\
 1 & 0.6666667 & 652 & UC \\
 157 & 4.5000000  &  0  &PA \\
 157 & 2.3333333  & 80  &PA \\
 157 & 0.8333333  & 171  &PA\\
 157 & 0.1666667 & 334 & PA\\
 \hline 
 \end{tabular}
 \caption{\footnotesize Data for the first participant in each treatment arm of the synthetic HAP2 data.}
 \label{tab_data}
 \end{table}

In this analysis, we assume that any participant with fewer than the maximum number of post-baseline visits  remains at risk for subsequent observations until the study end time, defined here as the maximum observed time across both UC and PA groups plus one day (1406 days). Thus, we set the argument \code{add.terminal.observations = TRUE} in the \code{fit\_SensIAT\_fulldata\_model()} function.

To match the model specification, we set the input parameters in \code{fit\_SensIAT\_fulldata\_model()} as follows:

\begin{example*}
model_fit <- fit_SensIAT_fulldata_model(
                data = HAP2_data,
                trt = Trt == "PA",
                id = "ID",
                outcome = "outcome",
                time = "time",
                alpha = c(-0.7, -0.5, -0.3, 0, 0.3, 0.5, 0.7),
                End  = 1406,
                intensity.args = list(bandwidth = 30),
                outcome_modeler = fit_SensIAT_single_index_fixed_coef_model, 
                outcome.args = list(abs.tol = 1e-7,
                                    kernel = "dnorm",
                                    model.modifications = ~. 
                                        - ns(..prev_outcome.., df = 3)
                                        - scale(..time..)
                                        - scale(..delta_time..)
                                        +..prev_outcome..+..time..+..delta_time..),
                knots = c(76, 654, 1232),
                add.terminal.observations = TRUE) 
\end{example*}

\noindent  Here \code{trt = Trt == "PA"} specifies that \code{Trt} is the column of the data frame \code{data} that contains the treatment assignment variable and that \code{PA} is the treatment group, while the next three arguments specify the names of the columns containing the participant identifier variable, the outcome variable, and the assessment time variable.  The argument \code{outcome\_modeler = fit\_SensIAT\_single\_index\_fixed\_coef\_}\-\code{model} specifies that the single index model for the observed outcomes distribution will be fit with the fixed intercept optimization method.  In \code{outcome.args}, \code{abs.tol} is the absolute tolerance parameter for optimization within the single index model; \code{kernel} is the kernel used for the single index model, set to Gaussian kernel; and \code{model} (or \code{model.modifications}) is the outcome model formula.  Here the predictors in the default outcome model formula are removed and  replaced with \code{..prev\_outcome..}, \code{..time..}, and \code{..delta\_time..}.\ The argument \code{knots} specifies the cubic B-splines $\boldsymbol{B}(t)$, with one interior knot at $t= 654$, and the \code{add.terminal.observations = TRUE} argument is as discussed above.

In the following analysis, we focus on the marginal mean functions for each arm, $\mu^{(a)}(t)$, $a=0,1$, and the difference in the marginal mean functions $\mu^{(1)}(t) - \mu^{(0)}(t)$ (i.e., treatment effect). For each estimate, we report the point estimate along with 95\% pointwise Wald confidence intervals, constructed using both the IF-based variance estimator and the jackknife variance estimator.

\begin{figure}[H]
\centering
\begin{subfigure}[b]{0.92\textwidth}
    \centering
    \includegraphics[width=\textwidth]
    {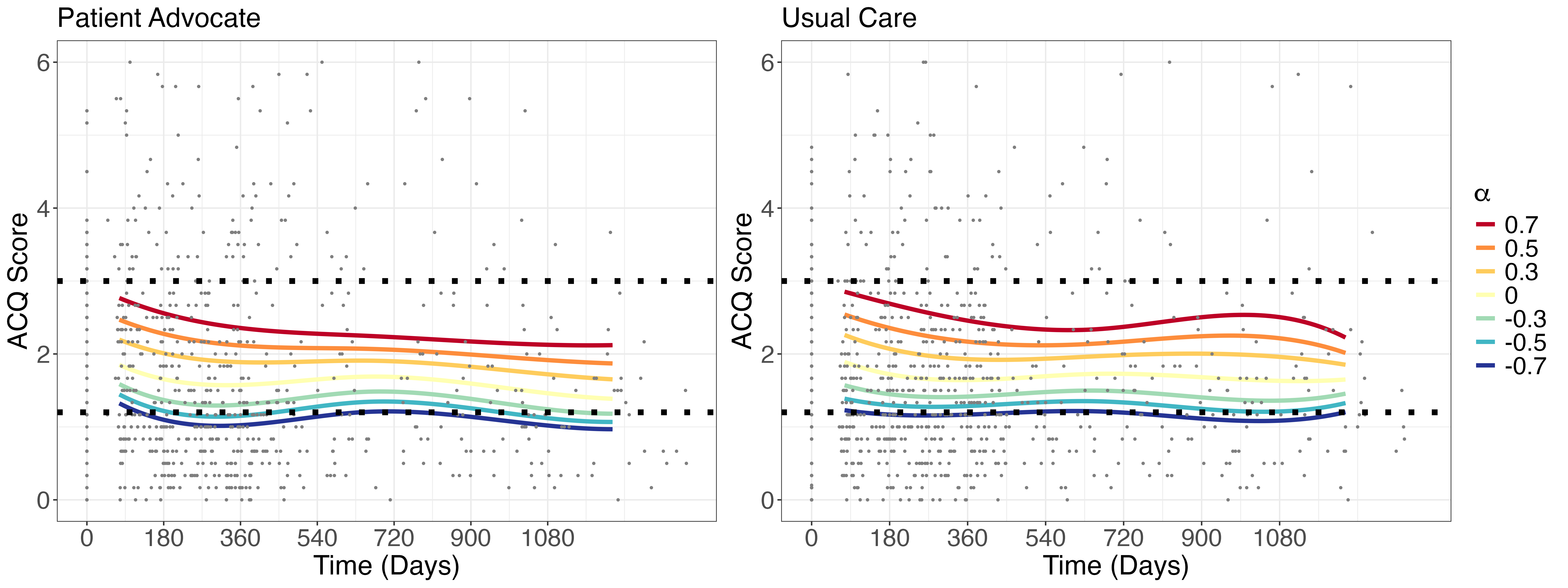}
\end{subfigure}
\vskip\baselineskip
\begin{subfigure}[b]{0.92\textwidth}
    \centering
    \includegraphics[width=\textwidth]
    {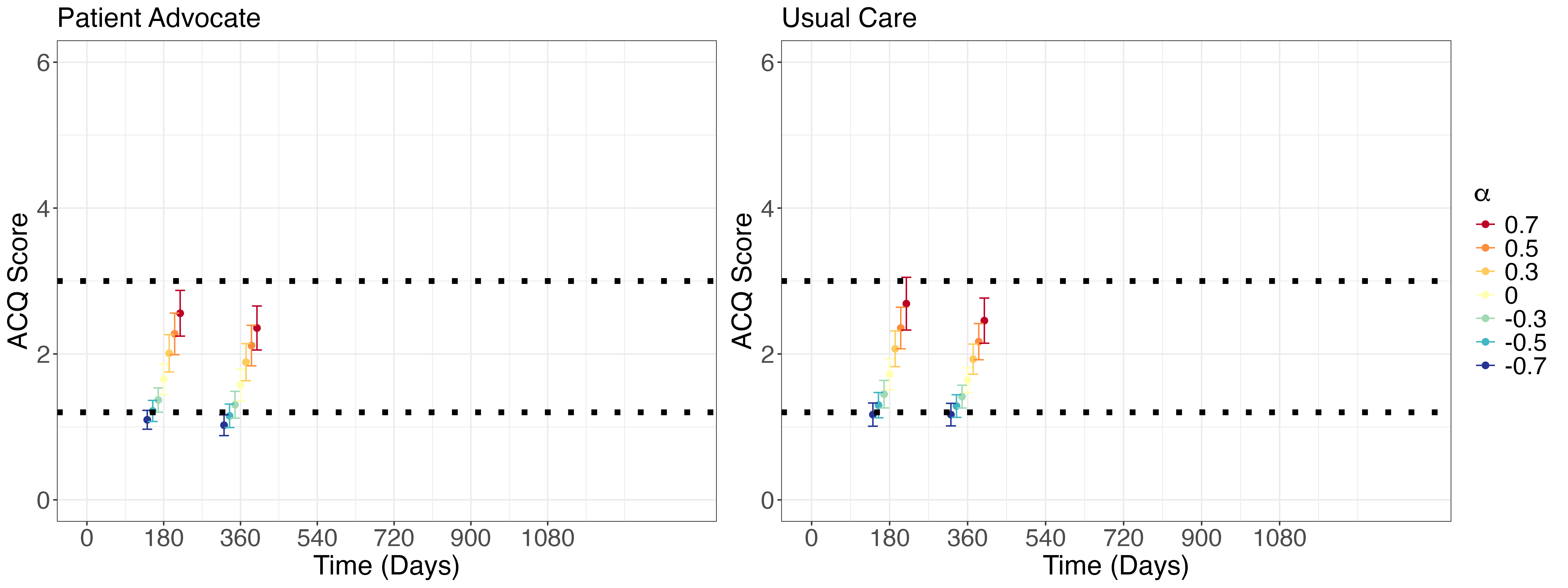}
\end{subfigure}
\caption{ \footnotesize 
Mean outcomes in the synthetic HAP2 data under a range of sensitivity parameter values. Here we show inference for the mean ACQ score in the treatment (patient advocate) and control (usual care) group, under values of $\alpha^{(0)}, \alpha^{(1)} = -0.7, -0.5, -0.3, 0, 0.3, 0.5, 0.7$. The upper panels show the estimated curves of mean ACQ score at time $t$, for $t = 76$ to $1232$ days after randomization. The lower panels show estimates, 95\% Wald confidence intervals using the jackknife variance estimate for the mean at time $t=180$ and $t=360$. For each group, only those values of $\alpha$ under which the mean falls between the dotted lines at $\mu_{min} = 1.2$ and $\mu_{max} = 3$ at all times are considered plausible based on expertise.}
\label{fig_fixinte:mean and CI}
\end{figure}

We first present the estimation results for the treatment-specific marginal mean function $\mu^{(a)}(t)$. Figure \ref{fig_fixinte:mean and CI} shows the observed values of $Y(t)$ (grey dots) and the estimated curves for $\mu^{(a)}(t)$ over the interval $76\leq t \leq 1232$ days (upper panels). The lower panels display point estimates and 95\% Wald confidence intervals (with jackknife standard errors) at the targeted time points. These plots were generated using the output from the \code{autoplot()} function as a base with additional enhancements such as layers for the original data and expert bounds. Specifically, the bases for the upper panels are obtained through the \code{autoplot} method for a \code{SensIAT\_within\_group\_model} object, which can be either the output from \code{fit\_SensIAT\_within\_group\_model()}, or one treatment arm within the output from  \code{fit\_SensIAT\_fulldata\_model()}.

\begin{example*}
base.PA <- autoplot(model_fit\$treatment)
base.UC <- autoplot(model_fit\$control)
\end{example*}

\noindent  To generate the plots in the lower panels, we first applied the \texttt{jackknife()} function to \code{model\_fit}.  Bases for the plots in the lower panels are obtained by applying the \code{autoplot} method to single-arm attributes of this jackknife object, separately for each treatment arm:
\begin{example*}
jack_all <- jackknife(model_fit, time = c(180, 360))
base.jk.PA <- autoplot(attr(jack_all, 'summary_treatment'))
base.jk.UC <- autoplot(attr(jack_all, 'summary_control'))
\end{example*}

According to our clinical collaborator, a mean ACQ score of 1.2 or lower, or 3 or higher, at any time point is considered clinically extreme. These thresholds are overlaid in Figure \ref{fig_fixinte:mean and CI}. As shown, for the PA group, sensitivity parameter values $\alpha^{(1)} = \{-0.7,-0.5,-0.3\}$ 
yield mean estimates that fall outside the plausible range. Similarly, for the UC group, $\alpha^{(0)} = -0.7$ yields mean estimates outside the plausible range.

Table \ref{tab_fixinte:mean_crl} and Table \ref{tab_fixinte:mean_trt} present the estimated mean ACQ scores at 6 and 12 months for the UC and PA groups, respectively. Results in Table \ref{tab_fixinte:mean_crl} and Table \ref{tab_fixinte:mean_trt} can be computed based on \code{jack\_all}. For the ACQ score, a lower value indicates better asthma control. As $\alpha^{(a)}$ increases, the estimated marginal mean also increases.

\renewcommand{\arraystretch}{1.5}

\begin{table}[H]

\footnotesize

\centering

 \begin{tabular}{|c |c || c |c |c |}
\hline
\multicolumn{5}{|c|}{Estimation results of the marginal mean in Usual Care group}\\
\hline
\multirow{2}{0.3in}{$\alpha^{(0)}$}&\multirow{2}{0.3in}{Visit}  &  \multicolumn{3}{c|}{Estimation results } \\
 \cline{3-5} 
 \ & \ & Est Mean & Wald CI (IF) & Wald CI (JK)\\
  \hline 
\multirow{2}{0.3in}{-0.5} 
& 6 months & 1.3 &$(1.17,1.42)$ & $(1.12,1.47)$ \\
 \cline{2-5} 
\ & 12 months & 1.29 &$(1.16, 1.41)$ & $(1.13, 1.44)$ \\
 \hline 
 \hline 
\multirow{2}{0.3in}{-0.3} 
& 6 months & 1.45 &$(1.31, 1.59)$ & $(1.26, 1.64)$ \\
 \cline{2-5} 
\ & 12 months & 1.42 &$(1.29, 1.55)$ & $(1.26, 1.57)$ \\
 \hline 
 \hline
 \multirow{2}{0.3in}{0} 
 & 6 months &1.72 & $(1.55, 1.89)$ & $(1.51, 1.94)$\\
 \cline{2-5} 
\ & 12 months&1.64 &$(1.49, 1.79)$& $(1.47, 1.81)$\\
 \hline 
 \hline
 \multirow{2}{0.3in}{0.3} 
 & 6 months & 2.07 &$(1.87, 2.27)$&$(1.83, 2.32)$\\
 \cline{2-5} 
\ & 12 months & 1.93 &$(1.75, 2.11)$& $(1.72, 2.14)$\\
 \hline 
 \hline
 \multirow{2}{0.3in}{0.5} 
 & 6 months & 2.36 & $(2.13, 2.59)$ & $(2.07, 2.64)$\\
 \cline{2-5} 
\ & 12 months & 2.17 & $(1.96, 2.38)$& $(1.92, 2.42)$ \\
 \hline 
 \hline
 \multirow{2}{0.3in}{0.7} 
 & 6 months & 2.69 &$(2.41, 2.97)$ & $(2.33, 3.05)$\\
 \cline{2-5} 
\ & 12 months & 2.46 &$(2.21, 2.70)$&$(2.15, 2.77)$\\
 \hline 
 \end{tabular}

  \caption{\footnotesize Sensitivity analysis for the marginal mean in the Usual Care group of the synthetic HAP2 data.  Shown are the point estimate, 95\% Wald confidence interval using the IF-based variance estimate, and 95\% Wald confidence interval using the jackknife variance estimate, under a selected range of $\alpha^{(0)}$ values.}

 \label{tab_fixinte:mean_crl}

 \end{table}

\renewcommand{\arraystretch}{1.5}

\begin{table}[H]

\footnotesize
\centering

 \begin{tabular}{|c |c || c |c |c |}
\hline
\multicolumn{5}{|c|}{Estimation results of the marginal mean in Patient Advocate group}\\
\hline
\multirow{2}{0.3in}{$\alpha^{(1)}$}&\multirow{2}{0.3in}{Visit}  &  \multicolumn{3}{c|}{Estimation results } \\
 \cline{3-5} 
 \ & \ & Est Mean & Wald CI (IF) & Wald CI (JK)\\
 \hline 
 \hline
 \multirow{2}{0.3in}{0} 
 & 6 months & 1.65 & $(1.48, 1.83)$&$(1.44, 1.86)$\\
 \cline{2-5} 
\ & 12 months& 1.57 & $(1.39, 1.76)$&$(1.36, 1.79)$\\
 \hline 
 \hline
 \multirow{2}{0.3in}{0.3} 
 & 6 months & 2.01 & $(1.79, 2.23)$& $(1.75, 2.26)$\\
 \cline{2-5} 
\ & 12 months & 1.89 & $(1.67, 2.1)$ & $(1.63, 2.14)$\\
 \hline 
 \hline
 \multirow{2}{0.3in}{0.5} 
 & 6 months & 2.28 & $(2.03, 2.52)$ & $(1.99, 2.56)$\\
 \cline{2-5} 
\ & 12 months & 2.11 & $(1.88, 2.35)$& $(1.84, 2.39)$ \\
 \hline 
 \hline
 \multirow{2}{0.3in}{0.7} 
 & 6 months & 2.56 & $(2.3, 2.82)$&$(2.24, 2.87)$\\
 \cline{2-5} 
\ & 12 months & 2.36 & $(2.11, 2.6)$&$(2.05, 2.66)$\\
 \hline 
 \end{tabular}
\caption{\footnotesize Sensitivity analysis for the marginal mean in the Patient Advocate group of the synthetic HAP2 data.  Shown are the point estimate, 95\% Wald confidence interval using the IF-based variance estimate, and 95\% Wald confidence interval using the jackknife variance estimate, under a selected range of $\alpha^{(1)}$ values.}
 
 \label{tab_fixinte:mean_trt}

 \end{table}

Finally, we evaluated the treatment effect at 6 and 12 months.

\renewcommand{\arraystretch}{1.5}

\renewcommand{\arraystretch}{2}

\begin{table}[H]

\fontsize{6pt}{6pt} \selectfont

\centering

\begin{tabular}{|p{0.07in} |p{0.2in} || p{0.48in} | p{0.53in} | p{0.63in} | p{0.66in} | p{0.66in} | p{0.66in} |}
\hline
\multicolumn{8}{|c|}{Inference for treatment effects at 6 and 12 months}\\
\hline
 \multirow{2}{0.07in}{}& \multirow{2}{0.19in}{6 months} &  \multicolumn{6}{c|}{$\alpha^{(0)}$} \\
\cline{3-8} 
 \ & \  & \multirow{1}{0.2in}{-0.5} & \multirow{1}{0.2in}{-0.3}  & \multirow{1}{0.2in}{0} & \multirow{1}{0.2in}{0.3} & \multirow{1}{0.2in}{0.5} & \multirow{1}{0.2in}{0.7}\\
 \hline 
 \multirow{4}{0.07in}{$\alpha^{(1)}$} 
 & 0.7 & \textcolor{orange}{$1.26 (0.9, 1.62)$} & \textcolor{orange}{$1.11 (0.74,1.48 )$}& \textcolor{orange}{$ 0.84(0.46, 1.21)$} & \textcolor{orange}{$0.49 (0.09, 0.88)$} & $0.2(-0.22,0.63)$ &  $-0.13(-0.61,0.35)$\\
 \cline{2-8} & 0.5 & \textcolor{orange}{$ 0.98 (0.64, 1.31)$} & \textcolor{orange}{$ 0.83 (0.49, 1.17)$} & \textcolor{orange}{$0.55 (0.2, 0.91)$} & $0.2(-0.17, 0.58)$ & $-0.08(-0.48,0.32)$ &$-0.41(-0.87,0.05)$\\
 \cline{2-8} & 0.3 & \textcolor{orange}{$0.71 (0.4,1.02)$}  & \textcolor{orange}{$0.56 (0.24, 0.88)$} & $0.29(-0.05, 0.62)$ & $-0.06(-0.42, 0.29)$ & $-0.35(-0.73, 0.04)$ & \textcolor{green}{$-0.68(-1.12,-0.24)$}\\
 \cline{2-8} & 0 & \textcolor{orange}{$0.36(0.08,0.63)$}&$0.21 (-0.08, 0.49)$ &$-0.07(-0.37, 0.23)$ & \textcolor{green}{$-0.42(-0.74, -0.09)$} & \textcolor{green}{$-0.7(-1.06, -0.35)$} & \textcolor{green}{$-1.03(-1.45,-0.62)$}\\
 \hline 
 \hline
 \multirow{2}{0.07in}{}&
 \multirow{2}{0.19in}{12 months}&
 \multicolumn{6}{c|}{$\alpha^{(0)}$ } \\
\cline{3-8} 
 &  & \multirow{1}{0.2in}{-0.5} & \multirow{1}{0.2in}{-0.3}  & \multirow{1}{0.2in}{0} & \multirow{1}{0.2in}{0.3} & \multirow{1}{0.2in}{0.5} & \multirow{1}{0.2in}{0.7}\\
 \hline 
 \multirow{4}{0.07in}{$\alpha^{(1)}$} & 0.7 & \textcolor{orange}{$1.07(0.73,1.41)$} & \textcolor{orange}{$0.94(0.6,1.28)$} & \textcolor{orange}{$0.71(0.37, 1.06)$} & \textcolor{orange}{$0.43(0.06,0.79)$} & $0.19 (-0.2,0.58)$ & $-0.1(-0.53,0.33)$\\
 \cline{2-8} 
 & 0.5 & \textcolor{orange}{$0.83 (0.51, 1.15)$} & \textcolor{orange}{$0.7 (0.38, 1.02)$} & \textcolor{orange}{$0.47 (0.15, 0.8)$} &$0.19(-0.16, 0.53)$ & $-0.05(-0.43,0.32)$ & $-0.34(-0.76,0.07)$\\
 \cline{2-8} 
& 0.3 &\textcolor{orange}{$0.6 (0.3, 0.9)$}& \textcolor{orange}{$0.47 (0.17, 0.77)$} &$0.25(-0.06, 0.55)$& $-0.04(-0.37, 0.29)$& $-0.28(-0.64,0.07)$ & \textcolor{green}{$-0.57(-0.97,-0.17)$}\\
\cline{2-8} 
 & 0 & \textcolor{orange}{$0.29(0.02,0.56)$} & $0.16(-0.11, 0.43)$&$-0.07(-0.34,0.21)$&\textcolor{green}{$-0.35(-0.66, -0.05)$}&\textcolor{green}{$-0.59(-0.93,-0.26)$}&\textcolor{green}{$-0.88(-1.26,-0.5)$}\\
 \hline 
 \end{tabular}

 \caption{ \footnotesize Sensitivity analysis for the treatment effect at 6 and 12 months
in the synthetic HAP2 data. For various choices of $\alpha^{(1)}$ and $\alpha^{(0)}$, we present estimates of treatment effects along with 95\% Wald confidence intervals (using jackknife standard errors). Entries in green correspond to values of $\alpha^{(1)}$ and $\alpha^{(0)}$ under which there would be evidence that the patient advocate reduces (that is, improves) the mean ACQ score, compared to the usual care. The entries in orange correspond to values of $\alpha^{(1)}$ and $\alpha^{(0)}$ under which there would be evidence that the patient advocate raises the mean ACQ score, compared to the usual care.}

\label{tab_fixinte:relative_diff}
 
\end{table}

Table \ref{tab_fixinte:relative_diff} presents the estimated  treatment effect 
and corresponding confidence intervals under selected values of $\alpha^{(0)}$ and $\alpha^{(1)}$, at 6 months and 12 months.
Assuming the explainable assessment assumption within each group ($\alpha^{(0)} = \alpha^{(1)}$ = 0), we find insufficient evidence to support a treatment effect
at both 6 months and 12 months. Similarly, under the assumption that the informativeness of assessments is the same across groups ($\alpha^{(0)} = \alpha^{(1)}$), there is insufficient evidence of a treatment effect at the targeted assessment times.
When allowing for differential informativeness between groups, there is evidence of a treatment effect in some cases but not in others. For example, if we assume explainable assessment in the PA group $(\alpha^{(1)} = 0)$ and informative assessment in the UC group ($\alpha^{(0)} \geq 0.3$; unobserved values are skewed higher), there is evidence that the PA intervention improves asthma control at both 6 and 12 months. Conversely, if we assume explainable assessment in the UC group $(\alpha^{(0)} = 0)$ and informative assessment in the PA group ($\alpha^{(1)} \geq 0.5$; unobserved values are skewed higher), there is evidence that the PA intervention results in worse asthma control relative to UC.

Next, we set the input parameter \code{alpha = seq(-0.7, 0.7, by = 0.02)} in \code{fit\_SensIAT\_fulldata\_}\-\code{model()}, while keeping all other input parameters unchanged, to obtain the result \code{model\_fit2} and \code{jack\_all2} on a finer grid of $\alpha^{(0)}$ and $\alpha^{(1)}$ values. Based on the result \code{model\_fit2}, we determined the most reasonable ranges of sensitivity parameter to be $-0.26 \leq \alpha^{(1)} \leq 0.7$ for the PA group and $-0.5 \leq \alpha^{(0)} \leq 0.7$ for the UC group.

\begin{figure}[H]
\centering
\begin{subfigure}[b]{0.95\textwidth}
    \centering
    \includegraphics[width=\textwidth]
    {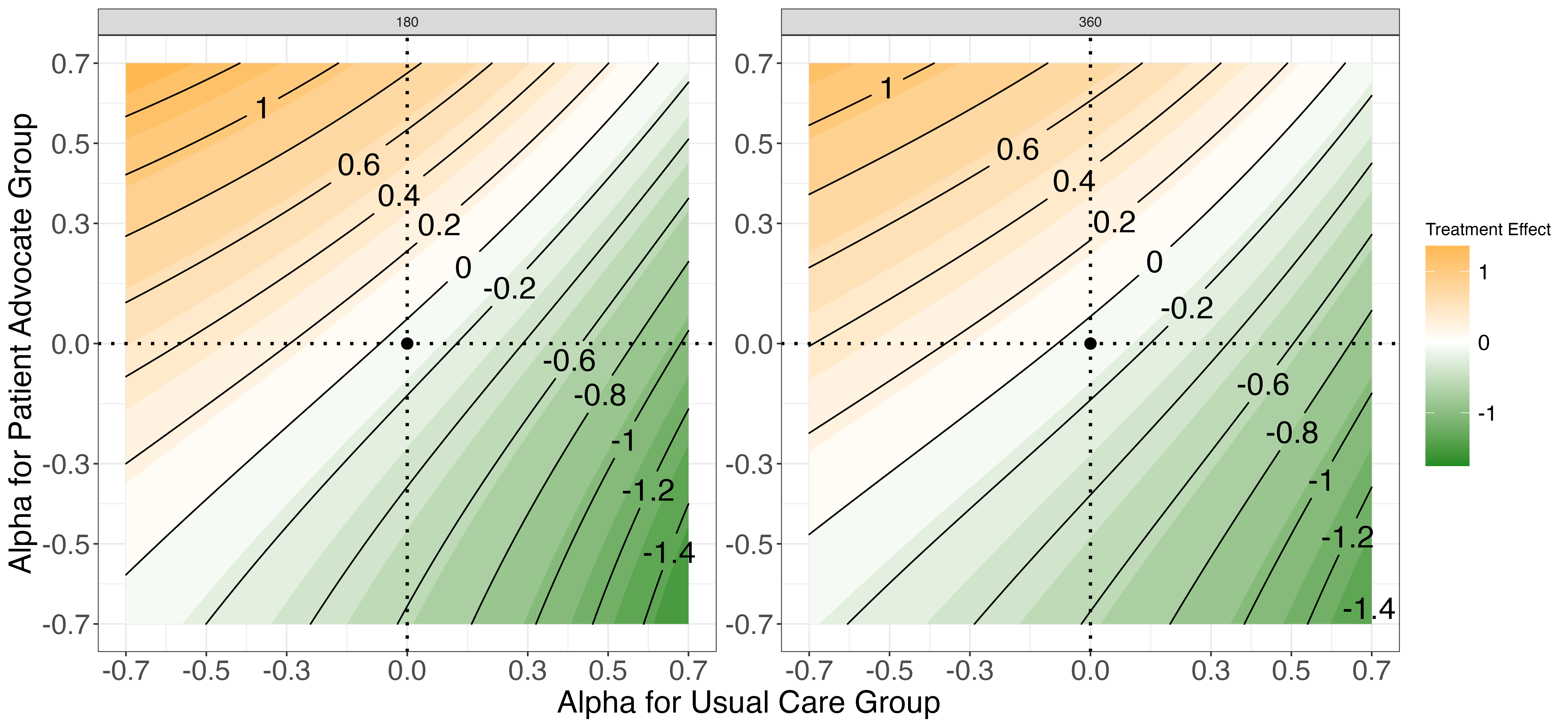}
    \caption{Point estimates at 6 months and 12 months}
\end{subfigure}
\vskip\baselineskip
\begin{subfigure}[b]{0.95\textwidth}   
    \centering 
    \includegraphics[width=\textwidth]{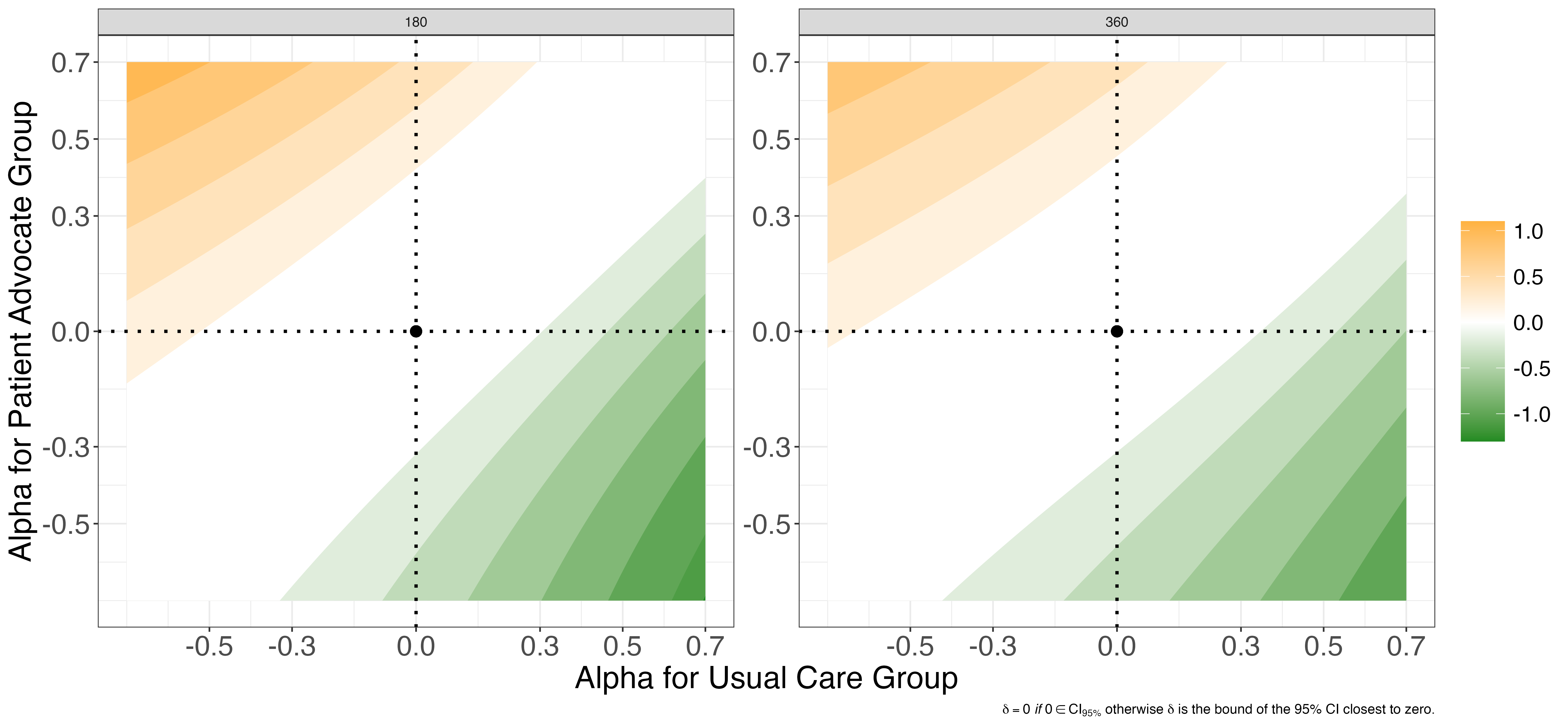}
    \caption{Confidence intervals information at 6 months and 12 months}
\end{subfigure}
\caption{\footnotesize
  Inferences over a range of sensitivity parameter values. Here, we present inference for 
  the treatment effect  at 6 months and 12 months, for a range of treatment-specific sensitivity parameter values.  Panels (a) shows the point estimates for the treatment effects for various combinations of  treatment-specific sensitivity parameter values. Panels (b) provide information about the confidence intervals: the white region corresponds to combinations of sensitivity parameter values for which the confidence interval includes zero. In the green regions, confidence intervals are entirely negative, and the values shown are the upper bound of the confidence interval. In the orange regions, confidence interval are entirely positive, and the values shown  are the lower bound of the confidence interval. Confidence intervals are Wald confidence intervals using jackknife standard errors.}
\label{fig_fixinte:contour}
\end{figure}

Figure~\ref{fig_fixinte:contour} displays a sensitivity analysis of the treatment effect 
at 6 and 12 months over a grid of $\alpha^{(0)}$ and $\alpha^{(1)}$ values. The horizontal and vertical axes represents $\alpha^{(0)}$ and $\alpha^{(1)}$ values, respectively. Panel (a) shows point estimates of the treatment effect.  This plot is obtained through the \code{autoplot} method applied to the \code{SensIAT\_fulldata\_model} object:

\begin{example*}
base.contour.point <- autoplot(model_fit2, time = c(180, 360))
\end{example*}
Panel (b) is generated from the output of the \code{jackknife()} performed on the full model:
\begin{example*}
base.contour.CI <- autoplot(jack_all2)
\end{example*}
In Panel (b), white regions indicate parameter values for which the confidence interval includes zero. Green regions represent values where the confidence interval is entirely negative, indicating that the PA intervention demonstrates better asthma control than UC. In contrast, orange regions correspond to intervals entirely above zero, favoring UC.

These results suggest that when $\alpha^{(0)}$ and $\alpha^{(1)}$ are equal, there is insufficient evidence to conclude a treatment difference at both 6 months and 12 months. However, for some cases when $\alpha^{(0)}>\alpha^{(1)}$ (lower right corner of the confidence intervals contour plot (b)), there is consistent evidence that the PA intervention provides better asthma control than UC at both 6 months and 12 months. Conversely, for some cases when $\alpha^{(0)}<\alpha^{(1)}$ (upper left corner of the confidence intervals contour plots (b)), there is evidence that UC provides better asthma control than the PA intervention.
This sensitivity analysis demonstrates that, after investigating the impact of informative assessment, there is insufficient evidence to conclude an asthma control benefit of PA in the synthetic data.

\section{Simulation study} \label{sec:sim_study}

We assessed the finite-sample performance of our estimation procedure in a realistic simulation study based on the HAP2 data.  Separately for each treatment arm, we simulated data with a sample size of $N=200$ in each arm using the data-generating procedure described in \citep{smith2024sa}, modified to the HAP2 data;  see \citep{smith2024sa} for details.  The HAP2 data were used to estimate, separately by treatment group, (1) the empirical distribution of the baseline outcome, (2) assessment time (stratified) intensity model with previous outcome as covariate, and (3) negative binomial observed outcome distribution model with previous outcome, time and time since previous assessment as covariates. These treatment-specific models were treated as the truth and used to simulate observed datasets.  For each treatment group, baseline outcomes were simulated from the empirical distribution; assessment times were simulated using Ogata's Thinning Algorithm \citep{ogata1981lewis} based on the assessment-time model; and outcomes at each assessment was simulated using the negative binomial distribution model.

True values of $\boldsymbol{\beta}$ were computed under values of $\alpha= -0.6,-0.3,0,0.3,0.6$ as described in \citep{smith2024sa}.  We analyzed the simulated data using our AIIW  estimation approach.  For each scenario, we evaluated our approach in terms of empirical bias and confidence interval coverage probabilities across 500 simulations. We considered Wald confidence intervals using the IF-based variance estimate and Wald confidence intervals using the jackknife variance estimate.

Simulation results for the marginal means in each treatment arm at 6 and 12 months, along with their true values, are presented in Tables \ref{tab_fbw:simu_crl} and \ref{tab_fbw:simu_trt}.
Bias in both arms is generally close to zero. Coverage of Wald confidence intervals using the jackknife variance estimate is close to the nominal level of $0.95$ in most scenarios. In contrast, Wald intervals based on the influence function variance tend to undercover, with coverage ranging from $0.878$ to $0.940$.

Results for treatment effects at 6 and 12 months across combinations of $\alpha^{(0)}$ and $\alpha^{(1)}$ are shown in Table \ref{tab_fbw:simu_effect}. To assess finite-sample performance, we first analyzed the simulated data using the true values of $\alpha^{(0)}$ and $\alpha^{(1)}$. We then evaluated the sensitivity of the results by repeating the analysis under the explainable assessment assumption ($\alpha^{(0)} = \alpha^{(1)} = 0$). When using the true values, the treatment effect estimates showed small bias (absolute bias less than or equal to $0.01$ in each case) and confidence interval coverage was close to the nominal $0.95$, ranging from $0.944$ to $0.976$. However, when incorrectly assuming $\alpha^{(0)} = \alpha^{(1)} = 0$, the estimates often exhibited substantial bias and poor coverage, especially when $\alpha^{(0)} \neq \alpha^{(1)}$. These results highlight the importance of accounting for informative assessment times and conducting sensitivity analyses under varying assumptions.

\renewcommand{\arraystretch}{1.5}

\begin{table}

\footnotesize

\centering

\begin{tabular}{| c | c || c | c | c | c | c | }
\hline
$\alpha^{(0)}$  & Parameter & True Value & Emp Mean & $|$Bias$|$  &  Wald (IF)  &  Wald (Jackknife) \\
\hline
\hline
\multirow{2}{0.3in}{ -0.6 } & $E\{Y(6)\}$ &
 1.202 & 1.203 & 0.001  & 0.902 & 0.956\\
 \cline{2-7} 
 \ & $E\{Y(12)\}$&
 1.143 & 1.146 &0.003 & 0.932 & 0.966\\
 \hline
\hline
\multirow{2}{0.3in}{ -0.3 }
 & $E\{Y(6)\}$ 
  & 1.430 &  1.427 &  0.003 & 0.912 &  0.954 \\
 \cline{2-7} 
 \ & $E\{Y(12)\}$ 
  & 1.352 & 1.353 & 0.001 &  0.924 &  0.958 \\
\hline
\hline
\multirow{2}{0.3in}{ 0 } 
& $E\{Y(6)\}$ 
   &1.721 & 1.716 & 0.005 & 0.898 &  0.956 \\
 \cline{2-7} 
 \ & $E\{Y(12)\}$  
  &1.617 & 1.616 & 0.001 & 0.922 & 0.962 \\
\hline
\hline
\multirow{2}{0.3in}{ 0.3 } 
& $E\{Y(6)\}$ 
   &2.078 & 2.073 & 0.005 & 0.904 &  0.952 \\
 \cline{2-7} 
 \ & $E\{Y(12)\}$  
  &1.943 & 1.943 & 0.000 & 0.926 & 0.960 \\
\hline
\hline
\multirow{2}{0.3in}{ 0.6 } 
& $E\{Y(6)\}$ 
   & 2.495 & 2.489 & 0.006 &  0.878 & 0.950 \\
 \cline{2-7}  
 \ & $E\{Y(12)\}$  
  & 2.330 &  2.332 & 0.002 &  0.898 &  0.976 \\
\hline
 \end{tabular}
 
 \smallskip
 
 \caption{\footnotesize Usual care arm simulation results (N = 200). Shown are the true values of the two target parameters under each of seven different data-generating mechanisms that mimic the usual care arm of the HAP2 data with $\alpha^{(0)}=-0.6,-0.3,0,0.3,0.6$; the empirical mean and absolute value of the empirical bias of the estimators across 500 simulations; and the coverage of Wald confidence intervals using IF-based variance estimate (Wald(IF)) and Wald confidence intervals using the jackknife variance estimate (Wald(Jackknife)). }

\label{tab_fbw:simu_crl}

 \end{table}


\begin{table}

\footnotesize

\centering

\begin{tabular}{| c | c || c | c | c | c | c | }
\hline
$\alpha^{(1)}$  & Parameter & True Value & Emp Mean & $|$Bias$|$  &  Wald (IF)  &  Wald (Jackknife) \\
\hline
\hline
\multirow{2}{0.3in}{ -0.6 } & $E[Y(6)]$ &
 1.215 & 1.221 & 0.006  & 0.920 & 0.964 \\
 \cline{2-7} 
 \ & $E[Y(12)]$&
 1.139 & 1.140 & 0.001 & 0.918 & 0.970\\
 \hline
\hline
\multirow{2}{0.3in}{ -0.3 }
 & $E[Y(6)]$ 
  & 1.449 &  1.452 &  0.003 & 0.930 &  0.962 \\
 \cline{2-7} 
 \ & $E[Y(12)]$ 
  & 1.350 & 1.348 & 0.002 &  0.934 &  0.972 \\
\hline
\hline
\multirow{2}{0.3in}{ 0 } 
& $E[Y(6)]$ 
   &1.747 & 1.748 & 0.001 & 0.928 &  0.964 \\
 \cline{2-7} 
 \ & $E[Y(12)]$  
  &1.617 & 1.614 & 0.003 & 0.940 & 0.968 \\
\hline
\hline
\multirow{2}{0.3in}{ 0.3 } 
& $E[Y(6)]$ 
   &2.112 & 2.112 & 0.000 & 0.922 &  0.950 \\
 \cline{2-7} 
 \ & $E[Y(12)]$  
  &1.946 & 1.945 & 0.001 & 0.920 & 0.962 \\
\hline
\hline
\multirow{2}{0.3in}{ 0.6 } 
& $E[Y(6)]$ 
   & 2.534 & 2.536 & 0.002 &  0.902 & 0.938 \\
 \cline{2-7}  
 \ & $E[Y(12)]$  
  & 2.337 &  2.341 & 0.004 &  0.884 &  0.954 \\
\hline
 \end{tabular}
 
 \smallskip
 
\caption{\footnotesize Patient advocate arm simulation results (N = 200). Shown are the true values of the two target parameters under each of seven different data-generating mechanisms that mimic the patient advocate arm of the HAP2 data with $\alpha^{(1)}=-0.6,-0.3,0,0.3,0.6$; the empirical mean and absolute value of the empirical bias of the estimators across 500 simulations; and the coverage of Wald confidence intervals using IF-based variance estimate (Wald(IF)) and Wald confidence intervals using the jackknife variance estimate (Wald(Jackknife)). }

\label{tab_fbw:simu_trt}

 \end{table}

\renewcommand{\arraystretch}{1.2}

\renewcommand{\arraystretch}{1.8}

\begin{table}

\fontsize{8pt}{8pt} \selectfont

\centering

\begin{tabular}{ c c c c  c  c  c  c c c c c c }
\hline 
& & & \multicolumn{10}{c}{True $\alpha^{(0)}$ } \\
 \cline{4-13} 
   \multicolumn{3}{c}{~} & \multicolumn{2}{c}{-0.6} & \multicolumn{2}{c}{-0.3} & \multicolumn{2}{c}{0} & \multicolumn{2}{c}{0.3} & \multicolumn{2}{c}{0.6} \\
 \cline{4-13} 
 \multicolumn{3}{c}{~}  & $|$Bias$|$ &  Cov.  &  $|$Bias$|$  & Cov. &  $|$Bias$|$ & Cov.  & $|$Bias$|$  & Cov.   & $|$Bias$|$  & Cov.   \\
 \hline
 \multirow{3}{0.4in}{\textbf{Month} \\ \ \ \  \textbf{6} } 
 &  
\multirow{2}{*}{ 0.6 } & S.A. & 0.001 & 0.954 & 0.004 & 0.958 & 0.007 &  0.948 &  0.007  & 0.954 &  0.008 &  0.960 \\
\cline{3-13}
& &  Expl.  & 1.300 & 0.000 & 1.072 & 0.004 &0.781  & 0.022 & 0.424 & 0.176 & 0.007 & 0.940 \\
\cline{2-13}
 \multirow{8}{0.4in}{True \\ \ \ \ $\alpha^{(1)}$ } & \multirow{2}{*}{0.3} 
 & S.A. & 0.001 & 0.946 &  0.003 &  0.952 & 0.005 & 0.948 & 0.006 & 0.948 & 0.007 &  0.950 \\
 \cline{3-13}
&  & Expl. & 0.878 & 0.016 & 0.650 & 0.042 & 0.359 & 0.294 & 0.002 & 0.940 & 0.415 & 0.190 \\
\cline{2-13}
& \multirow{2}{*}{0} 
& S.A. & 0.000 & 0.954 & 0.003 &  0.952 & 0.006 &  0.946 & 0.006 &  0.944 & 0.007 &  0.948 \\
\cline{3-13}
& & Expl. & 0.513 & 0.096 & 0.285 & 0.486 & 0.006 & 0.946 & 0.363 & 0.308 & 0.780 & 0.002 \\
\cline{2-13}
&\multirow{2}{*}{ -0.3} & S.A. & 0.002 &  0.968  & 0.006 & 0.958 & 0.008 &  0.946 &  0.008  &  0.946 &  0.010 &  0.950 \\
\cline{3-13}
& & Expl. & 0.215 & 0.642 & 0.013 & 0.946 & 0.304 & 0.452 & 0.661 & 0.010 & 1.078 & 0.000  \\
\cline{2-13}
& \multirow{2}{*}{ -0.6} 
& S.A. & 0.005 &  0.966 &  0.009 &  0.960 &  0.011 & 0.950 & 0.012 &  0.950 & 0.013 & 0.944 \\
\cline{3-13}
& & Expl. & 0.019 &  0.946 &  0.247 &  0.614 & 0.538 & 0.046 & 0.895  & 0.000 & 1.312 & 0.000  \\
\hline 
 \multirow{3}{0.4in}{\textbf{Month} \\ \ \ \ \textbf{12} } 
 &  
\multirow{2}{*}{ 0.6 } & S.A. & 0.000 & 0.954 & 0.003 & 0.956 & 0.005 &  0.956  &  0.004  & 0.958  &0.002  &  0.970 \\
\cline{3-13}
& & Expl. & 1.196 & 0.000 & 0.987 & 0.000 & 0.722 & 0.004 & 0.396 & 0.148 & 0.009 & 0.966 \\
\cline{2-13}
\multirow{8}{0.4in}{True \\ \ \ \ $\alpha^{(1)}$ } & \multirow{2}{*}{ 0.3 } 
& S.A. & 0.004 &  0.958 &  0.001 & 0.962 & 0.001 &  0.964 &  0.000 & 0.964 &0.002  & 0.968 \\
\cline{3-13}
& & Expl. & 0.805 & 0.000 & 0.596 & 0.016 & 0.331 & 0.272 & 0.005 & 0.968 & 0.382 & 0.162 \\
\cline{2-13}
& \multirow{2}{*}{ 0 } 
& S.A. & 0.006 & 0.974 & 0.004 & 0.968 & 0.002 & 0.968 & 0.003 & 0.976 & 0.005 & 0.966 \\
\cline{3-13}
& & Expl. & 0.476 & 0.072 & 0.267 & 0.444 & 0.002 & 0.968 & 0.324 & 0.324 & 0.711 & 0.000  \\
\cline{2-13}
& \multirow{2}{*}{ -0.3 } 
& S.A. & 0.005 & 0.972 &  0.003 &  0.962 & 0.001 & 0.964 & 0.002 & 0.970 & 0.004 & 0.968 \\
\cline{3-13}
& & Expl. & 0.209 & 0.630 & 0.000 & 0.970 & 0.265 & 0.502 & 0.591 & 0.006 & 0.978 &0.000  \\
\cline{2-13}
& \multirow{2}{*}{ -0.6 } 
& S.A. & 0.002 & 0.964 & 0.001 & 0.952 & 0.003 &  0.956 & 0.002 & 0.960 & 0.000 & 0.972 \\
\cline{3-13}
& & Expl. & 0.002 & 0.970 & 0.211 & 0.650 & 0.476  & 0.028  & 0.802 & 0.000 & 1.189 &0.000  \\
\hline
\end{tabular}

\caption{ \footnotesize
Simulation results.  Data were generated under the assumption of our sensitivity analysis framework using values of $\alpha^{(0)},\alpha^{(1)}=-0.6,-0.3,0,0.3,0.6$.  The treatment effects at 6 and 12 months were then estimated using augmented inverse intensity weighted estimators: (a) using the true values of $\alpha^{(0)},\alpha^{(1)}$ (rows denoted "S.A."), and (b) under the explainable assessment assumption that $\alpha^{(0)}=\alpha^{(1)}=0$ (rows denoted "Expl.").  Shown are the absolute value of the empirical bias and the confidence interval coverage across 500 simulations. Confidence intervals are Wald confidence intervals using the jackknife variance estimate.  
}

\label{tab_fbw:simu_effect}

\end{table}

\section{Conclusion}

In this paper, we have presented a software package \CRANpkg{SensIAT} for conducting sensitivity analysis of a two arm randomized trial with a continuous outcome that is irregularly collected during the course of follow-up. The estimation procedure is based on augmented inverse intensity weighting and depends on specification of a stratified Andersen-Gill intensity model and a single index observed outcome distribution.

There are many methodological extensions that will increase the utility of the package including: (1) binary and count outcomes, (2) informative drop-out, (3) weighting the influence function to reduce the impact of assessments that are far from the target times, (4) auxiliary time dependent covariates, and (5) prospective observational studies.


\section*{Acknowledgments}

Andrew Redd and Yujing Gao contributed equally and share first authorship.

Research reported in this article was funded through a Patient Centered Outcomes Research Institute (PCORI) Award ME-2021C3-24972. Andrew Redd, Yujing Gao, Andrea Apter and Daniel Scharfstein received funding from this award. The views and statements presented in this article are solely the responsibility of the authors and do not necessarily represent the views of the PCORI. 

The HAP2 study (PI: Andrea Apter) was funded by the National Institutes of Health/National Heart, Lung, and Blood Institute (grant no. R18 HL116285).

Bonnie Smith’s research was partially supported by National Institutes of Health grant EB029977 from the National Institute of Biomedical Imaging and Bioengineering.  

Ravi Varadhan would like to acknowledge funding support from the National Institutes of Health grant NCI CCSG P30 CA006973 from the National Cancer Institute.

\bibliography{article}

\begin{thebibliography}{26}
\providecommand{\natexlab}[1]{#1}
\providecommand{\url}[1]{\texttt{#1}}
\expandafter\ifx\csname urlstyle\endcsname\relax
  \providecommand{\doi}[1]{doi: #1}\else
  \providecommand{\doi}{doi: \begingroup \urlstyle{rm}\Url}\fi

\bibitem[Andersen and Gill(1982)]{andersengill1982}
P.~K. Andersen and R.~D. Gill.
\newblock Cox's regression model for counting processes: A large sample study.
\newblock \emph{The Annals of Statistics}, 10\penalty0 (4):\penalty0 1100 -- 1120, 1982.
\newblock \doi{10.1214/aos/1176345976}.
\newblock URL \url{https://doi.org/10.1214/aos/1176345976}.

\bibitem[Apter et~al.(2020)Apter, Perez, Han, Ndicu, Localio, Park, Mullen, Klusaritz, Rogers, Cidav, Bryant-Stephens, Bender, Reisine, and Morales]{apter2020patientadvocates}
A.~J. Apter, L.~Perez, X.~Han, G.~Ndicu, A.~Localio, H.~Park, A.~N. Mullen, H.~Klusaritz, M.~Rogers, Z.~Cidav, T.~Bryant-Stephens, B.~G. Bender, S.~T. Reisine, and K.~H. Morales.
\newblock Patient advocates for low-income adults with moderate to severe asthma: A randomized clinical trial.
\newblock \emph{The Journal of Allergy and Clinical Immunology: In Practice}, 8\penalty0 (10):\penalty0 3466--3473.e11, 2020.
\newblock ISSN 2213-2198.
\newblock \doi{https://doi.org/10.1016/j.jaip.2020.06.058}.
\newblock URL \url{https://www.sciencedirect.com/science/article/pii/S2213219820307017}.

\bibitem[Borchers(2023)]{pracma}
H.~W. Borchers.
\newblock \emph{pracma: Practical Numerical Math Functions}, 2023.
\newblock URL \url{https://CRAN.R-project.org/package=pracma}.
\newblock R package version 2.4.4.

\bibitem[Breslow(1972)]{breslow1972discussion}
N.~E. Breslow.
\newblock Discussion on professor cox's paper.
\newblock \emph{Journal of the Royal Statistical Society: Series B (Methodological)}, 34\penalty0 (2):\penalty0 216--217, 1972.
\newblock \doi{10.1111/j.2517-6161.1972.tb00900.x}.
\newblock URL \url{https://doi.org/10.1111/j.2517-6161.1972.tb00900.x}.

\bibitem[B\r{u}\v{z}kov\'{a} and Lumley(2007)]{buzkova2007dependent}
P.~B\r{u}\v{z}kov\'{a} and T.~Lumley.
\newblock Longitudinal data analysis for generalized linear models with follow-up dependent on outcome-related variables.
\newblock \emph{The Canadian Journal of Statistics / La Revue Canadienne de Statistique}, 35\penalty0 (4):\penalty0 485--500, 2007.
\newblock ISSN 03195724.
\newblock URL \url{http://www.jstor.org/stable/20445273}.

\bibitem[Bůžková and Lumley(2009)]{buzkova2009repeated}
P.~Bůžková and T.~Lumley.
\newblock Semiparametric modeling of repeated measurements under outcome-dependent follow-up.
\newblock \emph{Statistics in Medicine}, 28\penalty0 (6):\penalty0 987--1003, 2009.
\newblock \doi{https://doi.org/10.1002/sim.3496}.
\newblock URL \url{https://onlinelibrary.wiley.com/doi/abs/10.1002/sim.3496}.

\bibitem[Chiang and Huang(2012)]{CHIANG2012271}
C.-T. Chiang and M.-Y. Huang.
\newblock New estimation and inference procedures for a single-index conditional distribution model.
\newblock \emph{Journal of Multivariate Analysis}, 111:\penalty0 271--285, 2012.
\newblock ISSN 0047-259X.
\newblock \doi{https://doi.org/10.1016/j.jmva.2012.04.003}.
\newblock URL \url{https://www.sciencedirect.com/science/article/pii/S0047259X12000942}.

\bibitem[Cox(1975)]{cox1975partiallikelihood}
D.~Cox.
\newblock Partial likelihood.
\newblock \emph{Biometrika}, 62\penalty0 (2):\penalty0 269--276, 1975.
\newblock \doi{10.1093/biomet/62.2.269}.
\newblock URL \url{https://doi.org/10.1093/biomet/62.2.269}.

\bibitem[Cox(1972)]{cox1972regression}
D.~R. Cox.
\newblock Regression models and life-tables.
\newblock \emph{Journal of the Royal Statistical Society: Series B (Methodological)}, 34\penalty0 (2):\penalty0 187--202, 1972.
\newblock \doi{https://doi.org/10.1111/j.2517-6161.1972.tb00899.x}.
\newblock URL \url{https://rss.onlinelibrary.wiley.com/doi/abs/10.1111/j.2517-6161.1972.tb00899.x}.

\bibitem[Franks et~al.(2020)Franks, D’Amour, and Feller]{franks2020observationalsa}
A.~M. Franks, A.~D’Amour, and A.~Feller.
\newblock Flexible sensitivity analysis for observational studies without observable implications.
\newblock \emph{Journal of the American Statistical Association}, 115\penalty0 (532):\penalty0 1730--1746, 2020.
\newblock \doi{10.1080/01621459.2019.1604369}.
\newblock URL \url{https://doi.org/10.1080/01621459.2019.1604369}.

\bibitem[Lin et~al.(2004)Lin, Scharfstein, and Rosenheck]{lin2004dependent}
H.~Lin, D.~O. Scharfstein, and R.~A. Rosenheck.
\newblock Analysis of longitudinal data with irregular, outcome-dependent follow-up.
\newblock \emph{Journal of the Royal Statistical Society: Series B (Statistical Methodology)}, 66\penalty0 (3):\penalty0 791--813, 2004.
\newblock \doi{https://doi.org/10.1111/j.1467-9868.2004.b5543.x}.
\newblock URL \url{https://rss.onlinelibrary.wiley.com/doi/abs/10.1111/j.1467-9868.2004.b5543.x}.

\bibitem[Martin et~al.(2020)Martin, Raim, Huang, and Adragni]{ManifoldOptim}
S.~Martin, A.~M. Raim, W.~Huang, and K.~P. Adragni.
\newblock {ManifoldOptim}: An {R} interface to the {ROPTLIB} library for riemannian manifold optimization.
\newblock \emph{Journal of Statistical Software}, 93\penalty0 (1):\penalty0 1--32, 2020.
\newblock \doi{10.18637/jss.v093.i01}.

\bibitem[Naimi et~al.(2021)Naimi, Mishler, and Kennedy]{naimi2021machinelearning}
A.~I. Naimi, A.~E. Mishler, and E.~H. Kennedy.
\newblock Challenges in obtaining valid causal effect estimates with machine learning algorithms.
\newblock \emph{American Journal of Epidemiology}, 07 2021.
\newblock ISSN 0002-9262.
\newblock \doi{10.1093/aje/kwab201}.
\newblock URL \url{https://doi.org/10.1093/aje/kwab201}.
\newblock kwab201.

\bibitem[{National Research Council}(2010)]{nationalacademy2010missingdata}
{National Research Council}.
\newblock \emph{The Prevention and Treatment of Missing Data in Clinical Trials}.
\newblock The National Academies Press, Washington, DC, 2010.
\newblock ISBN 978-0-309-15814-5.
\newblock \doi{10.17226/12955}.
\newblock URL \url{https://www.nap.edu/catalog/12955/the-prevention-and-treatment-of-missing-data-in-clinical-trials}.

\bibitem[Ogata(1981)]{ogata1981lewis}
Y.~Ogata.
\newblock On lewis' simulation method for point processes.
\newblock \emph{IEEE Transactions on Information Theory}, 27\penalty0 (1):\penalty0 23--31, 1981.
\newblock \doi{10.1109/TIT.1981.1056305}.
\newblock URL \url{https://doi.org/10.1109/TIT.1981.1056305}.

\bibitem[Pullenayegum and Feldman(2013)]{pullenayegum2013dr}
E.~M. Pullenayegum and B.~M. Feldman.
\newblock Doubly robust estimation, optimally truncated inverse-intensity weighting and increment-based methods for the analysis of irregularly observed longitudinal data.
\newblock \emph{Statistics in Medicine}, 32\penalty0 (6):\penalty0 1054--1072, 2013.
\newblock \doi{10.1002/sim.5640}.
\newblock URL \url{https://onlinelibrary.wiley.com/doi/abs/10.1002/sim.5640}.

\bibitem[Pullenayegum and Lim(2016)]{pullenayegum2016review}
E.~M. Pullenayegum and L.~S. Lim.
\newblock Longitudinal data subject to irregular observation: A review of methods with a focus on visit processes, assumptions, and study design.
\newblock \emph{Statistical Methods in Medical Research}, 25\penalty0 (6):\penalty0 2992--3014, 2016.
\newblock \doi{10.1177/0962280214536537}.
\newblock URL \url{https://doi.org/10.1177/0962280214536537}.
\newblock PMID: 24855119.

\bibitem[Pullenayegum and Scharfstein(2022)]{pullenayegum2022repeatedly}
E.~M. Pullenayegum and D.~O. Scharfstein.
\newblock {Randomized trials with repeatedly measured outcomes: Handling irregular and potentially informative assessment times}.
\newblock \emph{Epidemiologic Reviews}, 44\penalty0 (1):\penalty0 121--137, 10 2022.
\newblock ISSN 1478-6729.
\newblock \doi{10.1093/epirev/mxac010}.
\newblock URL \url{https://doi.org/10.1093/epirev/mxac010}.

\bibitem[Redd(2012)]{redd2012comment}
A.~Redd.
\newblock A comment on the orthogonalization of b-spline basis functions and their derivatives.
\newblock \emph{Statistics and Computing}, 22\penalty0 (1):\penalty0 251--257, 2012.
\newblock \doi{10.1007/s11222-010-9221-0}.
\newblock URL \url{https://doi.org/10.1007/s11222-010-9221-0}.

\bibitem[Redd et~al.(2025)Redd, Gao, Yang, Smith, Mallett, and Scharfstein]{GHSensIAT}
A.~Redd, Y.~Gao, S.~Yang, B.~Smith, A.~Mallett, and D.~Scharfstein.
\newblock Sensiat: Sensitivity analysis for irregular assessment times.
\newblock \url{https://github.com/UofUEpiBio/SensIAT}, 2025.
\newblock URL \url{https://github.com/UofUEpiBio/SensIAT}.
\newblock R package version 0.0.0.9000.

\bibitem[Scharfstein et~al.(1999)Scharfstein, Rotnitzky, and Robins]{scharfstein1999nonignorable}
D.~O. Scharfstein, A.~Rotnitzky, and J.~M. Robins.
\newblock Adjusting for nonignorable drop-out using semiparametric nonresponse models.
\newblock \emph{Journal of the American Statistical Association}, 94\penalty0 (448):\penalty0 1096--1120, 1999.
\newblock \doi{10.1080/01621459.1999.10473862}.
\newblock URL \url{https://www.tandfonline.com/doi/abs/10.1080/01621459.1999.10473862}.

\bibitem[Smith et~al.(2024)Smith, Gao, Yang, Varadhan, Apter, and Scharfstein]{smith2024sa}
B.~B. Smith, Y.~Gao, S.~Yang, R.~Varadhan, A.~J. Apter, and D.~O. Scharfstein.
\newblock Semi-parametric sensitivity analysis for trials with irregular and informative assessment times.
\newblock \emph{Biometrics}, 80\penalty0 (4):\penalty0 ujae154, 12 2024.
\newblock ISSN 0006-341X.
\newblock \doi{10.1093/biomtc/ujae154}.
\newblock URL \url{https://doi.org/10.1093/biomtc/ujae154}.

\bibitem[Sun et~al.(2016)Sun, Peng, Manatunga, and Marcus]{sun2016quantile}
X.~Sun, L.~Peng, A.~Manatunga, and M.~Marcus.
\newblock Quantile regression analysis of censored longitudinal data with irregular outcome-dependent follow-up.
\newblock \emph{Biometrics}, 72\penalty0 (1):\penalty0 64--73, 2016.
\newblock \doi{https://doi.org/10.1111/biom.12367}.
\newblock URL \url{https://onlinelibrary.wiley.com/doi/abs/10.1111/biom.12367}.

\bibitem[Wang and Xia(2008)]{wang2008sliced}
H.~Wang and Y.~Xia.
\newblock Sliced regression for dimension reduction.
\newblock \emph{Journal of the American Statistical Association}, 103\penalty0 (482):\penalty0 811--821, 2008.
\newblock \doi{10.1198/016214508000000418}.
\newblock URL \url{https://doi.org/10.1198/016214508000000418}.

\bibitem[Wells(1994)]{wells1994kernelestimation}
M.~T. Wells.
\newblock {Nonparametric kernel estimation in counting processes with explanatory variables}.
\newblock \emph{Biometrika}, 81\penalty0 (4):\penalty0 795--801, 12 1994.
\newblock ISSN 0006-3444.
\newblock \doi{10.1093/biomet/81.4.795}.
\newblock URL \url{https://doi.org/10.1093/biomet/81.4.795}.

\bibitem[Xia et~al.(2002)Xia, Tong, Li, and Zhu]{xia2002adaptive}
Y.~Xia, H.~Tong, W.~K. Li, and L.-X. Zhu.
\newblock An adaptive estimation of dimension reduction space.
\newblock \emph{Journal of the Royal Statistical Society Series B: Statistical Methodology}, 64\penalty0 (3):\penalty0 363--410, 2002.
\newblock \doi{10.1111/1467-9868.03411}.
\newblock URL \url{https://doi.org/10.1111/1467-9868.03411}.

\end{thebibliography}

\address{Andrew Redd\\
  Division of Epidemiology\\
  Department of Internal Medicine\\
   University of Utah School of Medicine\\
  Salt Lake City, UT 84108\\
   USA\\
  \email{Andrew.Redd@hsc.utah.edu}}

\address{Yujing Gao \\
  Department of Statistics\\
  North Carolina State University\\
  Raleigh, NC 27695 \\
  USA\\
  \email{ygao39@ncsu.edu}}

\address{Bonnie Smith\\
  Department of Biostatistics\\
  Johns Hopkins Bloomberg School of Public Health \\
  Baltimore, MD 21205 \\
  USA\\
  \email{bsmit179@jhmi.edu}}

\address{Ravi Varadhan\\
  Department of Oncology \\
  Johns Hopkins School of Medicine\\
  Baltimore, MD 21205\\
  USA\\
  \email{ravi.varadhan@jhu.edu}}

  \address{Andrea Apter\\
  Pulmonary Allergy Critical Care Division\\
  Department of Medicine\\
  Perelman School of Medicine \\
  University of Pennsylvania\\
  Philadelphia, PA 19104\\
  USA \\
  \email{Andrea.Apter@pennmedicine.upenn.edu}}

  \address{Daniel Scharfstein\\
  Division of Biostatistics \\
  Department of Population Health Sciences\\
  University of Utah School of Medicine\\
  Salt Lake City, UT 84108\\
   USA\\
  \email{daniel.scharfstein@hsc.utah.edu}}

\end{article}

\end{document}